\documentclass[sigplan,10pt]{acmart}
\usepackage{xcolor}
\usepackage{xspace}
\usepackage{bm}
\usepackage{graphicx}
\usepackage{textcomp}
\usepackage{listings}
\usepackage{booktabs}   
\usepackage{pifont}    

\usepackage{mathtools}
\usepackage{amsthm}
\usepackage{multirow}
\usepackage{subfig}
\usepackage{tabularx}
\usepackage{rotating} 
\usepackage{algorithm} 
\usepackage[noend]{algpseudocode}
\usepackage{hyperref}
\hypersetup{hidelinks, pdfauthor={}, pdftitle={}, pdfsubject={}, pdfkeywords={}}

\theoremstyle{plain}

\theoremstyle{definition}

\theoremstyle{remark}

\definecolor{codegray}{gray}{0.95}
\definecolor{keywordblue}{rgb}{0.13,0.13,0.67}
\definecolor{commentgreen}{rgb}{0.0,0.45,0.0}
\definecolor{stringred}{rgb}{0.63,0.0,0.0}

\usepackage{tikz}
\usetikzlibrary{
    positioning,
    fit,
    calc,
    backgrounds,
    shadows,
    arrows.meta,
    shapes.geometric
}

\newcommand{\ourarch}{DFIT\xspace}
\newcommand{\cmark}{\ding{51}}  
\newcommand{\xmark}{\ding{55}}

\def\BibTeX{{\rm B\kern-.05em{\sc i\kern-.025em b}\kern-.08em
    T\kern-.1667em\lower.7ex\hbox{E}\kern-.125emX}}

\definecolor{usercolor}{RGB}{255,236,217}
\definecolor{userdraw}{RGB}{217,119,6}
\definecolor{corecolor}{RGB}{219,234,254}
\definecolor{coredraw}{RGB}{37,99,235}
\definecolor{viscolor}{RGB}{220,252,231}
\definecolor{visdraw}{RGB}{22,163,74}
\definecolor{opcolor}{RGB}{243,244,246}
\definecolor{opdraw}{RGB}{75,85,99}
\definecolor{normcolor}{RGB}{224,231,255}
\definecolor{normdraw}{RGB}{79,70,229}

\pgfdeclarelayer{background}
\pgfsetlayers{background,main}
\tikzset{
  basicbox/.style={
    rectangle,
    rounded corners=3pt,
    draw,
    thick,
    align=center,
    font=\sffamily\footnotesize,
    minimum width=2.4cm,
    minimum height=1.0cm
  },
  userbox/.style={basicbox, fill=usercolor, draw=userdraw},
  corebox/.style={basicbox, fill=corecolor, draw=coredraw},
  visbox/.style={basicbox, fill=viscolor, draw=visdraw},
  opbox/.style={basicbox, fill=opcolor, draw=opdraw},
  normbox/.style={
    basicbox,
    fill=normcolor,
    draw=normdraw,
    font=\sffamily\footnotesize\bfseries
  },
  decision/.style={
    diamond,
    draw=coredraw,
    fill=corecolor,
    thick,
    align=center,
    font=\sffamily\scriptsize,
    aspect=1.8,
    inner sep=1pt
  },
  container/.style={
    rectangle,
    rounded corners=6pt,
    draw=gray!60,
    fill=gray!5,
    dashed,
    thick,
    inner sep=12pt
  },
  flowarrow/.style={->, >=Stealth, thick, draw=gray!70!black},
  heavyarrow/.style={->, >=Stealth, line width=1.2pt, draw=coredraw},
  labeltext/.style={
    font=\sffamily\scriptsize,
    text=gray!80!black,
    align=center
  }
}

\AtBeginDocument{%
  \providecommand\BibTeX{{%
    Bib\TeX}}}

\acmBadgeR[https://www.acm.org/publications/policies/artifact-review-and-badging-current]{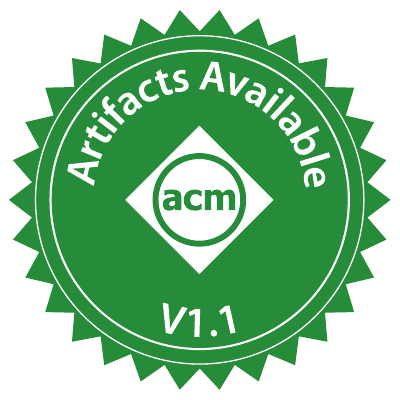}
\acmBadgeR[https://www.acm.org/publications/policies/artifact-review-and-badging-current]{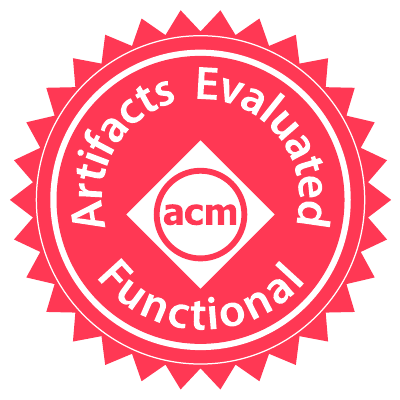}

\begin{document}

\title{\ourarch: A Framework for Building Instrumentation Tools Across Heterogeneous Dataflow Accelerators}
\author{Md. Azahar Alam}
\email{Md.Azahar.Alam@rockets.utoledo.edu}
\orcid{0009-0004-2072-7060}
\affiliation{
  \institution{The University of Toledo}
  \city{Toledo}
  \state{Ohio}
  \country{USA}
}
\author{Abrar Hossain}
\email{Abrar.Hossain@rockets.utoledo.edu}
\orcid{0009-0006-7007-8104}
\affiliation{
  \institution{The University of Toledo}
  \city{Toledo}
  \state{Ohio}
  \country{USA}
}

\author{Kishwar Ahmed}
\email{Kishwar.Ahmed@utoledo.edu}
\orcid{0000-0001-6295-3569}
\affiliation{
  \institution{The University of Toledo}
  \city{Toledo}
  \state{Ohio}
  \country{USA}
}








\renewcommand{\shortauthors}{Alam et al.}


\begin{abstract}

Emerging dataflow accelerators offer exceptional energy efficiency and throughput for machine learning workloads, yet remain largely opaque to developers for performance analysis. Unlike CPUs and GPUs,  where instrumentation frameworks enable fine-grained introspection, dataflow hardware exposes little information. This lack of transparency limits the ability of compiler engineers, system designers, and application developers to analyze, tune, or debug accelerator behavior.

In this paper, we introduce \textbf{\ourarch} (\underline{D}ata\underline{F}low Accelerator \underline{I}nstrumentation \underline{T}ool), an instrumentation framework designed specifically for dataflow architectures. \ourarch directly instruments operations, regions, and compiler transformations at the graph IR and backend-lowering stages using semantic program representation. \ourarch captures high-level, semantically meaningful data that reflect the true dataflow of computation. These data can be exported for visualization and analysis. Our approach demonstrates how instrumentation can be generalized beyond traditional hardware to support heterogeneous architectures, providing portability without modifying tool logic. 

By opening the black box of dataflow execution, \ourarch establishes a foundation for new classes of performance analysis tools for AI accelerators. It provides a practical path toward reproducible, architecture-aware introspection across heterogeneous systems, bridging the long-standing visibility gap between compiler design, runtime scheduling, and hardware realization. Evaluation across Cerebras, SambaNova, and TPU shows practical overhead of 3.53\%–13.62\% across instrumentation modes with robust cross-platform portability. 
 
\end{abstract}





\copyrightyear{2027}
\acmYear{2027}
\setcopyright{cc}
\setcctype{by}
\acmConference[EuroSys '27]{22nd European Conference on Computer Systems}{April 19--23, 2027}{Rabat, Morocco}
\acmBooktitle{22nd European Conference on Computer Systems (EuroSys '27), April 19--23, 2027, Rabat, Morocco}
\acmDOI{10.1145/3842654.3848581}
\acmISBN{979-8-4007-2971-3/2027/04}

\maketitle

 \section{Introduction}

Emerging machine learning accelerators, including TPUs~\cite{tpu-v1,tpu-data-center}, Cerebras Wafer Scale Engine~\cite{cerebras}, SambaNova~\cite{sambanova}, and other spatial dataflow architectures~\cite{eyeriss,leighton1986survey,siro2024}, achieve exceptional throughput by mapping tensor computations directly onto systolic arrays. Their execution model exploits parallel tensor pipelines, deterministic data movement, operator fusion, and backend-specific scheduling strategies to deliver substantial speedups for modern AI and scientific workloads. However, despite this performance advantage, these systems remain largely opaque to developers and compiler engineers. The only visibility typically provided comes through coarse aggregate metrics such as device utilization, kernel latency, or global memory  throughput~\cite{reddi2020mlperf,tflite-profiler,tensorflow}, which offer little insight into tensor movement, partition boundaries, or transformation effects during execution. As modern workloads grow more irregular, this absence of fine-grained introspection severely limits the ability to diagnose performance bottlenecks, verify correctness, or study microarchitectural behavior. As a result, the demand for transparent, low-overhead, semantically aligned instrumentation for dataflow accelerators has never been more pressing.

While dynamic instrumentation has a long history on CPUs and GPUs, existing approaches do not translate cleanly to MLIR-based~\cite{mlir} machine-learning compiler stacks~\cite{pytorch2} or to systolic-array accelerators. Frameworks such as Pin~\cite{pin}, DynamoRIO~\cite{dynamorio}, and Valgrind~\cite{valgrind} rely on sequential instruction streams, register-level semantics, and conventional control-flow graphs. GPU-oriented tools such as NVBit~\cite{nvbit}, CUPTI~\cite{cupti}, and SASSI~\cite{sassi,sassifi} assume SIMT execution models with thread blocks, warps, and instruction decoders that allow trampolines to be inserted around SASS instructions~\cite{sassi}. These approaches fundamentally depend on the existence of a well-defined instruction-level representation~\cite{nvidia-cuda-binary-utils,nvidia-cuda-driver-api,nvidia-cuda-fortran, nvidia-cuda-gdb,nvidia-nvcc,nvidia-ptx-isa,nvidia-cublas}, where instrumentation is attached to individual machine instructions or control-flow edges.

\begin{figure}[t]
\centering
\resizebox{\columnwidth}{!}{%
\begin{tikzpicture}[x=1cm,y=1cm]

\tikzset{
  lanehdr/.style={
    font=\sffamily\scriptsize\bfseries,
    text=gray!85!black,
    anchor=west
  },
  stagebox/.style={
    basicbox,
    minimum width=1.55cm,
    minimum height=0.78cm,
    font=\sffamily\scriptsize
  },
  binbox/.style={stagebox, fill=opcolor, draw=opdraw},
  dfbox/.style={stagebox, fill=viscolor, draw=visdraw},
  rtbox/.style={stagebox, fill=gray!8, draw=gray!60},
  hookbox/.style={
    rectangle,
    rounded corners=3pt,
    draw=opdraw,
    fill=opcolor,
    thick,
    align=center,
    font=\sffamily\tiny,
    minimum width=1.45cm,
    minimum height=0.55cm
  },
  smalltag/.style={
    rectangle,
    rounded corners=2pt,
    draw=normdraw,
    fill=normcolor,
    thick,
    align=center,
    font=\sffamily\tiny,
    minimum width=1.0cm,
    minimum height=0.42cm
  },
  flow/.style={->, >=Stealth, thick, draw=gray!70!black},
  hookarrow/.style={->, >=Stealth, line width=0.95pt, draw=opdraw},
  semarrow/.style={->, >=Stealth, line width=0.95pt, draw=normdraw},
  spanline/.style={line width=1.0pt, draw=normdraw},
  note/.style={font=\sffamily\tiny, text=gray!80!black, align=center}
}

\node[lanehdr] at (-0.6,1.55) {(a) Binary-Level Instrumentation};

\node[lanehdr] at (-0.6,-0.55) {(b) \ourarch Instrumentation};

\node[binbox] (bmodel)   at (0,0.75) {Model};
\node[binbox] (blower)   at (1.95,0.75) {Lowering};
\node[binbox] (bbin)     at (3.90,0.75) {Binary /\\SASS};
\node[rtbox]  (brun)     at (5.85,0.75) {Runtime};

\draw[flow] (bmodel.east) -- (blower.west);
\draw[flow] (blower.east) -- (bbin.west);
\draw[flow] (bbin.east)   -- (brun.west);

\node[note](binhook) at (3.90,-0.20) {Hook};
\draw[hookarrow] (binhook.north) -- (bbin.south);



\node[dfbox] (dmodel)  at (0,-1.5) {Model};
\node[dfbox] (dir)     at (1.95,-1.5) {IR\\Graph};
\node[dfbox] (dart)    at (3.90,-1.5) {Backend\\Lowering};
\node[dfbox] (drun)    at (5.85,-1.5) {Runtime};

\draw[flow] (dmodel.east) -- (dir.west);
\draw[flow] (dir.east) -- (dart.west);
\draw[flow] (dart.east) -- (drun.west);

\coordinate (spanL) at ($(dir.north west)+(-0.05,0.25)$);
\coordinate (spanR) at ($(dart.north east)+(0.05,0.25)$);

\draw[spanline] (spanL) -- (spanR);
\draw[spanline] (spanL) -- ++(0,-0.28);
\draw[spanline] (spanR) -- ++(0,-0.28);


\draw[semarrow] ($(spanL)!0.23!(spanR)$) -- (dir.north);
\draw[semarrow] ($(spanL)!0.77!(spanR)$) -- (dart.north);


\end{tikzpicture}%
}
\caption{(a) Binary-level runtime hooks; (b) \ourarch semantic instrumentation.}
\vspace{-0.3cm}
\label{fig:instr_compare_flow}
\end{figure}
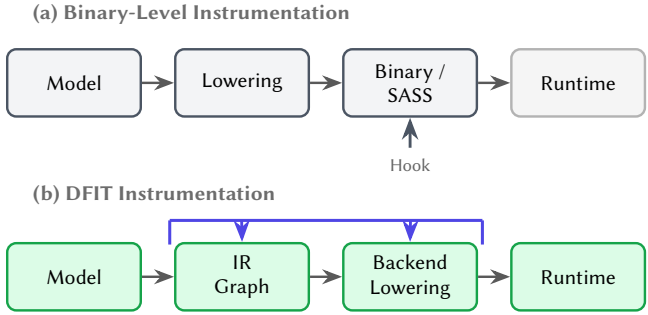

As a result, existing approaches force an undesirable trade-off: low-level instrumentation offers detailed runtime visibility but loses semantic meaning, while framework-level profilers preserve model semantics but provide little compiler or transformation awareness. This gap becomes even more severe for closed accelerator backends, where neither binary rewriting nor direct IR instrumentation is possible. As illustrated in Figure~\ref{fig:instr_compare_flow}, binary-level instrumentation hooks are introduced only after the program has been fully lowered into machine-level instructions, where higher-level operator structure and tensor relationships have already been erased. In contrast, the simplest we can do is to instrument earlier in the execution pipeline, spanning the Graph IR and backend-lowering stages where operator identity, tensor lineage, partition boundaries, and dataflow mappings~\cite{transformdialect2024} remain explicit and can be preserved for later runtime correlation.

Instead of depending on instruction-level rewriting, and to provide unified observability across accelerator stacks, we introduce a comprehensive framework, \textbf{\ourarch} (\underline{D}ata\underline{F}low Accelerator \underline{I}nstrumentation \underline{T}ool). This framework combines metadata propagation, runtime adapters, capability-aware backend interfaces, and normalized trace correlation to recover execution behavior in a form that remains useful for tool construction, portability analysis, and compiler-aware performance debugging. \ourarch is designed around four reusable services: a stable metadata and manifest layer, a capability-driven backend adapter interface, a policy-controlled trace pipeline, and a tool-facing API for developing customized analysis tools. This architecture enables the same tool logic to operate across Cerebras~\cite{cerebras}, SambaNova~\cite{sambanova}, and TPU while remaining robust to varying levels of backend capabilities.

This paper makes the following contributions:

\begin{itemize}
    \item We present \textbf{\ourarch}, an instrumentation framework for heterogeneous dataflow accelerators that enables instrumentation where no intermediate representation is exposed.
    
    \item We introduce a stable metadata model and normalized trace model that preserves operator identity, transformation lineage, tensor information, and execution boundaries across heterogeneous backends.
    
    \item We design a capability-driven adapter and policy architecture that enables portable instrumentation across closed compiler stacks and is demonstrated through TPU, Cerebras, and SambaNova.
    
    \item We demonstrate the practicality of \ourarch through (1) an Overhead Analysis Tool to demonstrate the overhead cost, scalability, and lightweightness, and (2) a Program Analyzer Tool to demonstrate the portability and extensibility.
\end{itemize}

 \section{Background}

\subsection{Dataflow Architecture}
\begin{figure}[t]
\centering
\resizebox{\columnwidth}{!}{
\begin{tikzpicture}[x=1cm,y=1cm]
\tikzset{
  pebox/.style={
    visbox,
    minimum width=0.55cm,
    minimum height=0.55cm,
    inner sep=0pt
  },
  membox/.style={
    opbox,
    minimum width=2.45cm,
    minimum height=0.95cm,
    font=\sffamily\scriptsize
  },
  ctrlbox/.style={
    corebox,
    minimum width=2.55cm,
    minimum height=0.95cm,
    font=\sffamily\scriptsize
  },
  outbox/.style={
    normbox,
    minimum width=2.3cm,
    minimum height=0.95cm,
    font=\sffamily\scriptsize
  },
  memflow/.style={->, >=Stealth, line width=1.0pt, draw=opdraw},
  ctrlflow/.style={->, >=Stealth, line width=1.0pt, draw=coredraw, dashed},
  peflow/.style={->, >=Stealth, line width=0.9pt, draw=visdraw},
  psumflow/.style={->, >=Stealth, line width=1.0pt, draw=normdraw},
  annot/.style={font=\sffamily\scriptsize, text=gray!80!black, align=center}
}

\node[userbox, minimum width=3.2cm, minimum height=1.05cm] (dram)
  at (-4.9, 7.4)
  {Host / Off-chip DRAM\\Weights $\mid$ Activations $\mid$ Outputs};

\node[corebox, minimum width=3.0cm, minimum height=0.95cm] (dma)
  at (-0.2, 7.4)
  {DMA / Transfer Engine};

\node[ctrlbox] (noc)
  at (-0.0, 5.8)
  {Tensor Router / NoC};

\node[ctrlbox] (sched)
  at (-3.5, 5.8)
  {Tile Controller\\\& Scheduler};

\node[membox] (wsram)
  at (-5.5, 3.5)
  {Weight SRAM};

\node[membox] (asram)
  at (0, 3.5)
  {Activation SRAM};


\node[pebox] (pe00) at (-4.8,  1.9) {};
\node[pebox] (pe01) at (-3.9,  1.9) {};
\node[pebox] (pe02) at (-3.0,  1.9) {};
\node[pebox] (pe03) at (-2.1,  1.9) {};

\node[pebox] (pe10) at (-4.8,  1.0) {};
\node[pebox] (pe11) at (-3.9,  1.0) {};
\node[pebox] (pe12) at (-3.0,  1.0) {};
\node[pebox] (pe13) at (-2.1,  1.0) {};

\node[pebox] (pe20) at (-4.8,  0.1) {};
\node[pebox] (pe21) at (-3.9,  0.1) {};
\node[pebox] (pe22) at (-3.0,  0.1) {};
\node[pebox] (pe23) at (-2.1,  0.1) {};

\node[pebox] (pe30) at (-4.8, -0.8) {};
\node[pebox] (pe31) at (-3.9, -0.8) {};
\node[pebox] (pe32) at (-3.0, -0.8) {};
\node[pebox] (pe33) at (-2.1, -0.8) {};

\node[
    rectangle,
    rounded corners=4pt,
    draw=red!60!black,
    fill=red!8,
    thick,
    minimum width=2.25cm,
    minimum height=2.45cm,
    anchor=center
] (pezoom) at ($(pe33.east)+(1.95,1.05)$) {};

\node[font=\sffamily\tiny\bfseries]
    at ($(pezoom.north)+(0,-0.18cm)$)
    {PE};

\node[
    rectangle,
    draw=red!50!black,
    fill=white,
    minimum width=0.8cm,
    minimum height=0.28cm,
    font=\sffamily\tiny
] (infifo) at ($(pezoom.center)+(-0.45,0.62)$) {In};

\node[
    rectangle,
    draw=red!50!black,
    fill=white,
    minimum width=0.8cm,
    minimum height=0.28cm,
    font=\sffamily\tiny
] (wfifo) at ($(pezoom.center)+(0.45,0.62)$) {Wt};

\node[
    rectangle,
    draw=red!60!black,
    fill=red!15,
    minimum width=0.75cm,
    minimum height=0.38cm,
    font=\sffamily\tiny\bfseries
] (mac) at ($(pezoom.center)+(0,0.1)$) {MAC};

\node[
    rectangle,
    draw=red!50!black,
    fill=white,
    minimum width=1.0cm,
    minimum height=0.28cm,
    font=\sffamily\tiny
] (psum) at ($(pezoom.center)+(0,-0.45)$) {Psum};

\node[
    rectangle,
    draw=red!50!black,
    fill=white,
    minimum width=0.75cm,
    minimum height=0.24cm,
    font=\sffamily\tiny
] (ctrl) at ($(pezoom.center)+(0,-0.95)$) {Comm Ports};

\draw[->, thin] (infifo.south) -- (mac.north west);
\draw[->, thin] (wfifo.south) -- (mac.north east);
\draw[->, thin] (mac.south) -- (psum.north);
\draw[->, thin] (psum.south) -- (ctrl.north);

\node[outbox] (reduce)
  at (-5.8, -2.5)
  {Output Reduc/\\Tree Merge/\\Normalize};

\node[outbox] (outsram)
  at (-3.1, -2.5)
  {Output SRAM\\Writeback\\FIFO};

\node[outbox] (hostout)
  at (-0.3, -2.5)
  {Host-visible\\Output\\Feature Maps/\\ Tensor Tiles};

\begin{pgfonlayer}{background}

  \coordinate (wsramfit) at ([xshift=-40pt]wsram.north);

  \node[container,
        fit=(wsramfit)(asram)(pe00)(pe33),
        inner sep=10pt] (midplane) {};
  \node[font=\sffamily\bfseries\scriptsize, text=gray!80!black,
        above=2pt of midplane.north] {Data Staging \& Compute};

  \node[container, fit=(dram)(dma)(noc)(sched), inner sep=8pt] (topplane) {};
  \node[font=\sffamily\bfseries\scriptsize, text=gray!80!black,
        above=2pt of topplane.north] {Control \& Routing};

  \node[container, fit=(pe00)(pe33), inner sep=8pt] (arraybox) {};

  \node[container,
        minimum width=8.6cm,
        minimum height=1.5cm,
        anchor=north] (outplane)
        at ($(arraybox.south)+(0.4,-0.32cm)$) {};
  \node[font=\sffamily\bfseries\scriptsize, text=gray!80!black,
        below=2pt of outplane.south] {Output Path};

\end{pgfonlayer}
\draw[flowarrow] (dram.east) -- (dma.west);
\draw[flowarrow] ([xshift=5pt]dma.south) -- ([xshift=0pt]noc.north);
\draw[ctrlflow] (sched.east) -- ++(0,0) |- (noc.west);
\draw[ctrlflow] (sched.south) |- (arraybox.north);

\draw[memflow]
(dma.south)
-- ++(0,-0.4)
-| (-1.5,3.68)
-- ([yshift=5pt]wsram.east);

\draw[memflow] (noc.south) -- (asram.north);

\draw[memflow] (wsram.south) -- ++(0,0) |- (pe00.west);
\draw[memflow] (wsram.south) -- ++(0,0) |- (pe10.west);
\draw[memflow] (wsram.south) -- ++(0,0) |- (pe20.west);
\draw[memflow] (wsram.south) -- ++(0,0) |- (pe30.west);

\node[annot] at (-5.9,1.25) {Weight\\stream};

\draw[memflow] (asram.south) -- ++(0,-0.35) -| (pe00.north);
\draw[memflow] (asram.south) -- ++(0,-0.35) -| (pe01.north);
\draw[memflow] (asram.south) -- ++(0,-0.35) -| (pe02.north);
\draw[memflow] (asram.south) -- ++(0,-0.35) -| (pe03.north);

\node[annot] at (-2.2,2.85) {Activation stream};

\foreach \a/\b in {
  pe00/pe01, pe01/pe02, pe02/pe03,
  pe10/pe11, pe11/pe12, pe12/pe13,
  pe20/pe21, pe21/pe22, pe22/pe23,
  pe30/pe31, pe31/pe32, pe32/pe33}
{
  \draw[peflow] (\a.east) -- (\b.west);
}

\foreach \a/\b in {
  pe00/pe10, pe10/pe20, pe20/pe30,
  pe01/pe11, pe11/pe21, pe21/pe31,
  pe02/pe12, pe12/pe22, pe22/pe32,
  pe03/pe13, pe13/pe23, pe23/pe33}
{
  \draw[peflow] (\a.south) -- (\b.north);
}

\foreach \a/\b in {
  pe00/pe11, pe01/pe12, pe02/pe13,
  pe10/pe21, pe11/pe22, pe12/pe23,
  pe20/pe31, pe21/pe32, pe22/pe33}
{
  \draw[ctrlflow] (\a.south east) -- (\b.north west);
}

\foreach \a/\b in {
  pe01/pe10, pe02/pe11, pe03/pe12,
  pe11/pe20, pe12/pe21, pe13/pe22,
  pe21/pe30, pe22/pe31, pe23/pe32}
{
  \draw[ctrlflow] (\a.south west) -- (\b.north east);
}

\draw[dashed, thick, red!60!black](pe33.north east) -- (pezoom.north west);
\draw[dashed, thick, red!60!black](pe33.south east) -- (pezoom.south west);

\draw[psumflow] (pe30.south) |- ($(reduce.north)+(0,0.52cm)$) -- (reduce.north);
\draw[psumflow] (pe31.south) |- ($(reduce.north)+(0,0.52cm)$) -- (reduce.north);
\draw[psumflow] (pe32.south) |- ($(reduce.north)+(0,0.52cm)$) -- (reduce.north);
\draw[psumflow] (pe33.south) |- ($(reduce.north)+(0,0.52cm)$) -- (reduce.north);

\draw[flowarrow] (reduce.east) -- (outsram.west);
\draw[flowarrow] (outsram.east) -- (hostout.west);






\end{tikzpicture}%
}
\caption{Unified dataflow architecture across modern AI accelerators.}
\vspace{-0.45cm}
\label{fig:dataflow_architecture}
\end{figure}
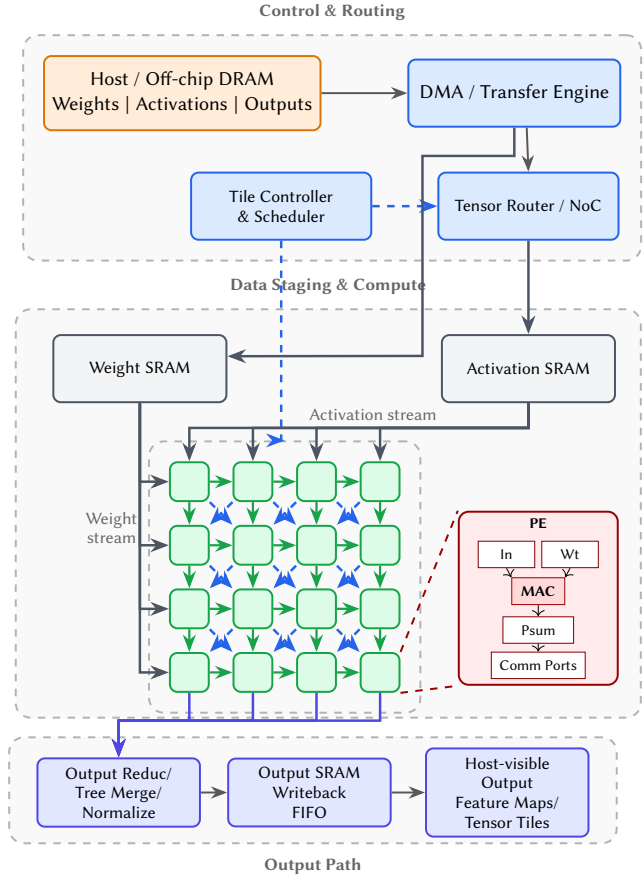

Modern machine learning accelerators such as Google's Tensor Processing Unit (TPU)~\cite{tpu-url},  Neural Processing Units (NPUs)~\cite{npu,mlirair2025}, Cerebras's Wafer Scale Engine~\cite{cerebras} are designed around a dataflow execution model, which differs fundamentally from the control-flow paradigm of CPUs and the SIMT model of GPUs~\cite{siro2024}. Instead of issuing sequences of instructions from a centralized control unit, a dataflow accelerator triggers computation when operands become available, as shown in Figure~\ref{fig:dataflow_architecture}. Each Processing Element (PE) performs a fixed micro-operation, typically a multiply–accumulate (MAC), and forwards results to neighboring elements through spatial interconnects~\cite{tpu-data-center,tpu-v1,tpu-url}. This enables massive parallelism and high utilization for dense linear algebra workloads.

At the architectural level~\cite{tpu-v1,cerebras,sambanova}, these accelerators employ large systolic or spatial dataflow fabrics, commonly organized as two-dimensional grids of PEs interconnected through short, regular data paths, as illustrated in Figure~\ref{fig:dataflow_architecture}. Each PE receives streams of operands, typically activations and weights, performs a fixed micro-operation such as multiply and accumulate, and forwards partial results to neighboring elements in a pipelined manner. The figure highlights this shared organization through SRAM-backed operand staging buffers, a spatial PE fabric, and a dedicated output path for reduction and writeback, which together capture the common execution structure across modern dataflow accelerators. Although the three platforms differ in realization, they share the same core architectural principle: execution is driven by operand movement and local propagation rather than by a centralized instruction stream. TPUs emphasize regular systolic execution for dense tensor workloads, Cerebras extends the spatial model to a wafer-scale fabric with distributed on-chip resources, and SambaNova adopts a more reconfigurable spatial dataflow organization. Consequently, the performance of these systems is governed primarily by the efficiency of operand propagation, tiling, and pipeline synchronization rather than by conventional instruction throughput~\cite{maestro,timeloop,scalesim}.

Unlike binary instrumentation for CPUs or GPUs~\cite{nvbit,pin, caliper, atom, sassi,sassifi}, where program analysis tools can instrument individual instructions or kernel launches, dataflow accelerators execute highly optimized firmware-controlled kernels~\cite{tpu-data-center} whose internal operations are not directly visible to developers. The microcode sequences that manage PE scheduling, tensor tiling, and activation streaming are proprietary and often opaque. This lack of observability makes it difficult to determine how efficiently PEs are utilized, how synchronization barriers affect throughput, or how different layers of a neural network interact during pipeline execution. Existing vendor profilers expose only aggregate utilization and latency statistics, offering little insight into runtime behavior~\cite{reddi2020mlperf, tflite-profiler}.

\subsection{MLIR Overview}
\subsubsection{Compiler Abstraction}
\label{sec:bg_compiler_gap}

A defining characteristic of ML and dataflow accelerator stacks is the progressive transformation of computation across multiple compiler abstraction levels. High-level graph operators may first be lowered into dialect-specific tensor operations~\cite{mlir,transformdialect2024}, then transformed through scheduling, fusion, partitioning, and memory-layout optimization passes before finally reaching a backend runtime representation. Each stage introduces performance-critical decisions while simultaneously moving the execution farther away from the original semantic source~\cite{lucke2025mlir}.

Traditional instrumentation systems are poorly aligned with this compiler structure because they typically assume a single stable observation layer. However, in multi-level compiler infrastructures such as MLIR~\cite{mlir}, semantic value is intentionally preserved across abstraction levels for as long as it remains useful. This compiler philosophy creates an opportunity for a fundamentally different form of instrumentation: rather than observing only the final execution form, the instrumentation framework can preserve stable identities, transformation lineage, and stage-aware metadata across lowering boundaries. This compiler-aware view is essential for understanding how optimization decisions manifest as runtime effects.

\subsubsection{Semantic Preservation}
\label{sec:bg_semantic_preservation}

The central systems challenge is therefore preserving semantic correspondence across transformations that intentionally rewrite, fuse, reorder, or partition the computation. A single high-level operator may expand into multiple backend partitions, while multiple source operators may be fused into one execution region~\cite{transformdialect2024}. Without a stable semantic contract, runtime events become difficult to attribute to source-level intent, compiler transformations, or tensor-level behavior~\cite{mlir-gpu-code-gen,hir}.

Addressing this challenge is a central design goal of \ourarch. Semantic preservation must survive both direct IR transformations and lowering passes, and must be reconstructable through canonical metadata manifests, graph artifacts, and boundary-level execution correlation. Treating semantic identity as a first-class instrumentation concern enables sharing a unified trace model, which later supports transformation attribution and cross-backend comparison.

\subsection{Instrumentation Fundamentals}

Dynamic Binary Instrumentation (DBI) provides a mechanism for observing, analyzing, and modifying a program's execution without requiring access to source code or recompilation. Unlike static analysis, which inspects binaries offline, DBI systems interpose on execution and inject user-defined analysis routines directly into the running program. This enables fine-grained inspection of events such as instruction execution, memory accesses, synchronization points, and architectural state transitions, while preserving the functional correctness of the original program.

At a high level, building an instrumentation tool on any conventional platform, CPU or GPU, requires three foundational capabilities. First, the instrumentation system must be able to \emph{observe} the execution stream, typically through decoding or intercepting the instructions that the processor executes. Second, it must be able to \emph{modify} this stream, either through binary rewriting, trampoline insertion, or dynamic translation, enabling callbacks to be executed before or after certain instructions or control-flow edges. Third, it must maintain a \emph{runtime mapping} between original program locations and their instrumented counterparts, often through a code cache or translation table~\cite{valgrind, dynamorio,pin}, to ensure that instrumented instructions execute correctly and efficiently.

\ourarch addresses these requirements by treating semantic identity, backend portability, and low overhead as first-class design constraints, as detailed in Section~\ref{sec:design_challenges}.
  \section{Architecture and Framework Design}
\label{sec:framework}

\subsection{Design Challenges and Key Decisions}
\label{sec:design_challenges}

Building an instrumentation framework for heterogeneous dataflow accelerators raises three systems challenges that do not arise in conventional binary instrumentation. 

\textbf{C1: Semantic identity across compilation boundaries.} Unlike CPUs and GPUs, where a stable instruction stream provides natural instrumentation anchors, dataflow accelerators progressively transform and partition operators across multiple compiler stages. A naive approach of attaching metadata at a single stage loses correspondence as operators are fused, tiled, or reordered. We address this by propagating stable semantic identifiers from the IR level through all lowering stages, so that runtime events can always be mapped back to their source operators regardless of backend transformations.

\textbf{C2: Heterogeneous backend visibility.} Backends differ fundamentally in what they expose. None of the three evaluated platforms provides a compiler-visible pass interface: Cerebras emits a dump of compiler artifacts, the Edge~TPU emits a compiler report and a model file, and SambaNova exposes runtime logs and sidecar artifacts. A single instrumentation mode cannot cover these. We address this through capability discovery at session initialization, which selects between compiler-visible semantic tagging and artifact reconstruction without changing the tool-facing API.

\textbf{C3: Tool portability without lowest common denominator APIs.} Portable frameworks often achieve portability by restricting tools to the weakest available abstraction. We instead normalize all backend observations into a common semantic event schema at the framework level, preserving the richest available information from each backend while exposing a single stable contract to tool developers.

\subsection{Architecture Overview}
\label{sec:arch_overview}
\begin{figure*}[!t]
\resizebox{\textwidth}{!}{%
\begin{tikzpicture}

    \node[userbox] (cli)      at (0, 1.4)  {dfabitctl\\(CLI Driver)};
    \node[userbox] (registry) at (0, 0)    {Tool\\Registry};
    \node[userbox] (session)  at (0, -1.4) {Session\\Management};

    \node[corebox] (api)     at (3.8, 0)    {\textbf{Tool Dev.}\\ \textbf{API Layer}\\(Unified)};
    \node[corebox] (context) at (3.8, -1.4) {Shared Context\\\& Services};

    \node[decision] (cap) at (7.2, 0) {Discover\\Capabilities()};

    \node[opbox]  (import)   at (10.0,  0) {Metadata and\\ Config Import};
    \node[visbox] (irtag)    at (12.6, 0) {IR Semantic\\Tagging};
    \node[visbox] (manifest) at (15.2, 0) {Manifest\\Export};

    \node[normbox, text width=2.2cm, minimum height=2.8cm, align=center]
    (norm) at (18.6, 0)
    {Normalization \&\\
     Event Correlation\\[4pt]
     \textit{(Normalized\\Trace Output)}};

    \begin{pgfonlayer}{background}
        \coordinate (topY) at (0,  2.4);
        \coordinate (botY) at (0, -2.4);

        \coordinate (z1L) at (-1.4, 0);
        \coordinate (z1R) at ( 1.4, 0);
        \node[container, fit=(z1L|-topY)(z1R|-botY)] (zone1) {};
        \node[font=\sffamily\bfseries\scriptsize, text=gray!80!black,
              above=4pt of zone1.north] {User Input Layer};

        \coordinate (z2L) at (2.8, 0);
        \coordinate (z2R) at (4.8, 0);
        \node[container, fit=(z2L|-topY)(z2R|-botY)] (zone2) {};
        \node[font=\sffamily\bfseries\scriptsize, text=gray!80!black,
              above=4pt of zone2.north] {Session \& Semantic Context};

        \coordinate (z3L) at (6.0,  0);
        \coordinate (z3R) at (16.4, 0);
        \node[container, fit=(z3L|-topY)(z3R|-botY)] (zone3) {};
        \node[font=\sffamily\bfseries\scriptsize, text=gray!80!black,above=4pt of zone3.north] {Capability-Driven Backend Adapter Layer};

    \end{pgfonlayer}


    \draw[flowarrow] (cli.east)      -- ++(0.4,0) |- (api.west);
    \draw[flowarrow] (registry.east) --            (api.west);
    \draw[flowarrow, dashed] (session.east) --     (context.west);

    \draw[heavyarrow] (api.east) -- (cap.west);

    \draw[heavyarrow] (cap.east)      -- (import.west);
    \draw[heavyarrow] (import.east)   -- (irtag.west);
    \draw[heavyarrow] (irtag.east)    -- (manifest.west);

    \draw[heavyarrow] (manifest.east) -- (norm.west);

\end{tikzpicture}
}
\caption{High-level execution flow of a \ourarch tool. After capability
discovery, the pipeline progresses through graph import, IR semantic tagging,
and manifest export before converging into a normalized semantic event stream,
allowing the tool to remain unchanged across heterogeneous backends.}
\label{fig:arch_overview}
\end{figure*}
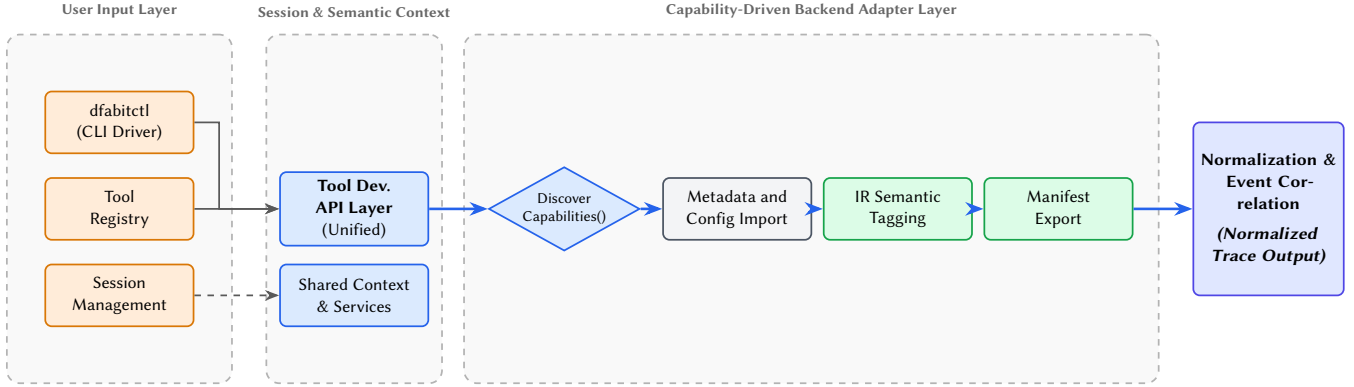

\ourarch addresses the challenges above through a layered architecture that separates semantic acquisition, event normalization, and tool execution. Figure~\ref{fig:arch_overview} shows the end-to-end flow. To illustrate concretely through \ourarch, we use a compact \emph{Compile Metrics} tool as the running example. The tool observes compile-stage metric samples, accumulates them through the standard tool lifecycle, and emits a normalized CSV summary at the end of the session. Although intentionally simple, this example exercises the same end-to-end workflow used by larger built-in tools, shown in Section \ref{sec:eval_port_ext}, Section \ref{sec:analysis}, and Section \ref{sec:use_cases}, while remaining small enough to be cleanly implemented.

\begin{lstlisting}[float,language=C++,basicstyle=\ttfamily\footnotesize,caption={Minimal compile-stage analysis tool implemented using the \ourarch lifecycle hooks.},label={lst:compile_metrics_tool}]
class CompileMetricsTool final : public Tool {
public:
    // Registry-visible tool name used by DFIT.
    std::string name() const override { return "compile_metrics"; }

    // Clear tool-local state before the session begins.
    core::Status OnInit(api::Context* ctx) override {
        if (!ctx) return {core::StatusCode::kInvalidArgument, "ctx is null"};
        collected_metrics_.clear();
        return core::Status::Ok();
    }

    // Consume compile-stage metric samples after compilation finishes.
    core::Status OnCompileEnd(
        api::Context* ctx,
        const adapters::CompileArtifactSet& ca) override {
        if (!ctx) return {core::StatusCode::kInvalidArgument, "ctx is null"};
        for (const auto& metric : ca.metrics)
            collected_metrics_.push_back(metric);
        return core::Status::Ok();
    }

    // Export the final normalized metric table when the session closes.
    core::Status OnShutdown(api::Context* ctx) override {
        if (!ctx) return {core::StatusCode::kInvalidArgument, "ctx is null"};
        return ToolServices::WriteMetricsCsv(
            "tool_compile_metrics.csv", collected_metrics_);
    }

private:
    std::vector<adapters::MetricSample> collected_metrics_;
};
\end{lstlisting}

Figure~\ref{fig:arch_overview}, together with Listing~\ref{lst:compile_metrics_tool}, illustrates how \ourarch executes through the framework. The end-to-end workflow is initiated through the control driver, \ourarch, which serves as the user-facing entry point of the framework. It resolves the selected backend adapter, instantiates the requested tool from the registry using the identifier returned by \texttt{name()}, and binds both into a single instrumentation session.

Once the session begins, the framework invokes \texttt{OnInit()} to reset tool-local state before any compile, load, or execution activity starts. From this point onward, the framework intercepts lifecycle events through the internal APIs, invoking the appropriate tool hooks as compilation, loading, and execution proceed. Upon detecting semantic identities across transformation stages, \ourarch's semantic correspondence generates compile reports and runtime boundary observations using \texttt{OnCompileEnd()}.

After execution completes and the session closes, OnShutdown() emits the final normalized CSV report using the shared reusable service layer. This stable lifecycle-driven workflow is what allows \ourarch tools to remain portable across backends with very different levels of compiler and runtime visibility.

The cost of adopting \ourarch is bounded by this interface. A tool author implements the lifecycle hooks and nothing further, defining behavior only for the stages the analysis needs and returning immediately from the rest. Listing~\ref{lst:compile_metrics_tool} does real work in three of them. The five built-in tools of Section~\ref{sec:use_cases} and Section~\ref{sec:eval_port_ext} range from 186 to 260 lines and follow the same shape. The knowledge required is C++ and the analysis itself. A tool author does not read compiler internals, vendor SDKs, or artifact formats, because the adapter normalizes those into operator descriptors and metric samples before any hook runs. The same tool source runs against every registered backend, and what differs between them is how much each descriptor carries, which the tool reads from the capability descriptor.

\subsection{Tool Development APIs}
\label{sec:tool_interface}

\ourarch exposes its tool-development interface as four API families that separate lifecycle control, semantic services, artifact-reconstruction services, and reusable helpers. This organization allows users to reason about the framework in terms of execution stages and semantic visibility rather than backend implementation details.

\subsubsection{Tool Lifecycle Hooks}
\label{sec:lifecycle_hooks}

Every tool in \ourarch implements the same lifecycle APIs. These hooks define when the framework transfers control to the tool during registration, initialization, compilation, loading, execution, and shutdown. For a tool developer, this API family forms the primary control surface of the framework.

\begin{lstlisting}[float,language=C++,basicstyle=\ttfamily\footnotesize,caption={Tool lifecycle APIs.},label={lst:tool_lifecycle_hooks}]
  // Lifecycle hooks invoked by DFIT across tool registration, session initialization, compile/load/run stages, and final shutdown.
  virtual std::string name() const = 0;
  virtual core::Status OnRegister(api::Context* ctx)
  virtual core::Status OnInit(api::Context* ctx)
  virtual core::Status OnCompileBegin(api::Context* ctx,const adapters::CompileArtifactSet& ca)
  virtual core::Status OnCompileEnd(api::Context* ctx,const adapters::CompileArtifactSet& ca)
  virtual core::Status OnLoadBegin(api::Context* ctx,const adapters::RuntimeArtifactSet& ra)
  virtual core::Status OnLoadEnd(api::Context* ctx,const adapters::RuntimeArtifactSet& ra)
  virtual core::Status OnRunBegin(api::Context* ctx,const adapters::RuntimeArtifactSet& ra)
  virtual core::Status OnRunEnd(api::Context* ctx,const adapters::RuntimeArtifactSet& ra)
  virtual core::Status OnShutdown(api::Context* ctx)

\end{lstlisting}




The lifecycle hooks in Listing~\ref{lst:tool_lifecycle_hooks} mirror the end-to-end workflow introduced in Section~\ref{sec:arch_overview}. \texttt{OnRegister()} binds the tool to the registry, allowing it to be resolved by the control driver. The compile, load, and run hooks allow tools to observe distinct execution stages without embedding backend-specific logic. This stage-aware contract is particularly important for tools that need to compare semantic fidelity before and after compilation or measure how much information survives partitioning and deployment.

\subsubsection{Semantic APIs}
\label{sec:mlir_services}

\ourarch exposes a richer semantic service layer that provides direct access to graph structure, stable semantic identities, transformation lineage, and compile-stage metadata.

\begin{lstlisting}[float,language=C++,basicstyle=\ttfamily\footnotesize,caption={Representative semantic APIs.},label={lst:mlir_mode_services}]
// Compiler visibility services includes Load, Build, Tag, export, and populate.
virtual core::Status LoadModule(const std::string& input)
virtual core::Status BuildModel(metadata::ModelDesc* model)
virtual core::Status TagStableIds(metadata::ModelDesc* model)
virtual core::Status ExportManifest(const metadata::ModelDesc& model)
virtual void PopulateStaticCost(metadata::OpDesc* op)
\end{lstlisting}
The services shown in Listing~\ref{lst:mlir_mode_services} allow a tool to reason directly about operators, tensors, subgraphs, and transformation stages without reconstructing semantic relationships from runtime traces. In practice, this mode is used by tools that require transformation attribution, compiler-stage correctness checks, static cost propagation, or operator-level memory analysis. Since semantic identities are preserved directly from the compiler-visible representation, the resulting trace provides the highest fidelity view available in the framework.

\subsubsection{Artifact-Reconstruction Services}
\label{sec:hidden_services}

Moreover, \ourarch provides artifact reconstruction services. Using these APIs, semantic meaning is rebuilt from the last visible graph artifact together with runtime boundary observations.

\begin{lstlisting}[float,language=C++,basicstyle=\ttfamily\footnotesize,caption={Representative artifact reconstruction services.},label={lst:hidden_mode_services},]
// Artifact reconstruction services for import, partition, tracking, and correlation.
virtual core::Status ImportGraph(const std::string& input)
virtual core::Status LoadSidecar(const std::string& input)
virtual core::Status BeginPartition(const std::string& name)
virtual core::Status EndPartition(const std::string& name)
virtual core::Status CorrelateRuntime(const std::vector<RuntimeRecord>& records)
virtual std::vector<adapters::MetricSample> GenerateMetricSamples()
\end{lstlisting}
The goal of the API family presented in Listing~\ref{lst:hidden_mode_services} is to preserve enough semantic structure for the tool to continue operating over a stable abstraction. Graph import and semantic recovery, rebuild node identity, tensor metadata, and partition layout, while runtime reconstruction reconnects these entities with execution boundaries and timing records. This design allows the same tool logic to remain valid even when the backend only exposes runtime artifacts and profiler summaries.

\subsubsection{Shared Context and Reusable Services}
\label{sec:shared_services}

The final API family contains the context objects and reusable services that remain identical across instrumentation modes. These helpers provide filtered metadata views, policy construction, metric export, and lightweight state sharing between execution stages.

\begin{lstlisting}[float,language=C++,basicstyle=\ttfamily\footnotesize,caption={Shared context and reusable services.},label={lst:shared_context_services}]
// Shared context and reusable services in DFIT are metadata, run config, framework config, property getters as well as filters and policy apis and writer apis.
virtual const std::vector<metadata::OpDesc>& GetMetadataView()
virtual const core::RunConfig& GetRunConfig()
virtual const core::FrameworkConfig& GetFrameworkConfig()
virtual std::string GetProperty(const std::string& key)
virtual void SetProperty(const std::string& key,std::string value)
virtual policy::PolicyEngine BuildPolicyEngine()
virtual std::vector<metadata::OpDesc> FilterOps()
virtual std::vector<adapters::MetricSample> FilterMetrics()
virtual core::Status WriteMetricsCsv(const std::string& path)
virtual core::Status WriteOpTable(const std::string& path)
\end{lstlisting}

The role of the API family in Listing~\ref{lst:shared_context_services} is the main portability layer for users. A tool can retrieve filtered metadata, store intermediate state between lifecycle stages, build custom tracing policies, and export structured results without depending on how semantic information was originally obtained. As a result, the dual-mode design affects only the semantic acquisition layer, while the surrounding tool logic, state management, and reporting pipeline remain unchanged.

\subsection{Software Components}
\label{sec:software_components}

The workflow above is implemented through a set of software components that correspond directly to the stages shown in Figure~\ref{fig:arch_overview}.

The common runtime core consists of the \emph{session manager}, \emph{tool registry}, \emph{policy controller}, \emph{trace sink}, and \emph{event normalizer}. These components are shared among instrumentation modes. The session manager controls the lifecycle of a tracing session, while the tool registry binds the selected analysis module to the normalized trace stream. The event normalizer guarantees that both instrumentation paths converge into the same normalized semantic event schema, which is the key portability contract exposed to tools~\cite{caliper,hpctoolkit}. Each backend contributes its own adapter implementation to this component stack, allowing the same session and normalization pipeline to remain unchanged while backend-specific artifact collection and runtime hooks vary underneath. At the control plane, \ourarch orchestrates backend selection, tool registration, session execution, and result export, providing a single user-facing entry point to the full software stack.

\ourarch builds on the toolchains rather than replacing any part of them. It compiles no models, schedules no kernels, and links no vendor SDK. Each backend already emits a record of what it compiled, and \ourarch reads that record: the CIRH MLIR the Cerebras compiler writes~\cite{mlir}, the summary the Edge~TPU compiler prints, and the model file the converter produced. Where a session drives compilation or execution, \ourarch invokes the vendor toolchain unmodified and collects what it leaves behind. What \ourarch implements is everything above that boundary. The stable identity scheme, the operator and tensor metadata model, the normalized event schema, the policy-controlled trace pipeline, and the tool API are framework code, and none of them exists in the stacks underneath. This division is what allows the same tool to run on a backend that exposes a compiler IR and one that exposes only a text report.

\subsection{Backend Adapter}
\label{sec:backend_capability}
The adapter layer is \ourarch's answer to Challenge C2: it isolates backend-specific artifact collection behind a uniform lifecycle contract, so that the capability discovery mechanism in Section~\ref{sec:design_challenges} can select the appropriate instrumentation mode transparently. Backend portability in \ourarch is realized through a uniform adapter contract that separates backend-specific artifact collection and runtime hooks from the mode-independent tool interface described in Section~\ref{sec:tool_interface}. Rather than exposing backend-specific compiler or instrumentation APIs directly to tools, each backend implements the same lifecycle-driven adapter interface~\cite{mlir,llvm_passinfra,caliper} and reports its semantic and runtime visibility through a capability descriptor. This contract allows the framework to preserve a stable execution workflow even when the backend exposes very different levels of compiler access.

\begin{lstlisting}[float,language=C++,basicstyle=\ttfamily\footnotesize,caption={Capability-driven backend integration flow.},label={lst:backend_capability_model}]
// Create backend adapter and initialize session state.
adapter = BackendRegistry::Create(selected_backend)
adapter.InitializeSession(ctx)

// Query backend visibility and runtime capabilities.
cap = adapter.DiscoverCapabilities()

// Select instrumentation mode from capability descriptor.
if (cap.visible_mlir || cap.visible_graph_ir)
    mode = COMPILER_VISIBLE
else
    mode = ARTIFACT_RECONSTRUCTION

// Execute backend lifecycle through the unified adapter contract.
adapter.PrepareArtifacts(ctx)
adapter.CompileBegin(ctx)
adapter.CompileEnd(ctx)
adapter.LoadBegin(ctx)
adapter.LoadEnd(ctx)
adapter.RunBegin(ctx)
adapter.RunEnd(ctx)

// Normalize runtime metrics and execution boundaries.
metrics = adapter.CollectRuntimeMetrics()
normalizer.Emit(metrics, mode)

// Close backend session.
adapter.Shutdown(ctx)
\end{lstlisting}

Listing~\ref{lst:backend_capability_model} explains how the framework unifies different backends into a single execution path using capability discovery and adapter creation. All backends follow a common lifecycle—covering preparation, compilation, execution, and metric collection—ensuring consistent higher-level tool behavior. This design enables easy extensibility, where new accelerators only need to implement adapter and capability logic, as shown by the Cerebras and SambaNova adapters evaluated in Section~\ref{sec:eval_port_ext}.

\subsubsection{Supporting a New Backend}
\label{sec:new_backend}

A new backend must expose a graph description in some readable form, name
operators consistently enough that identifiers assigned before lowering still
resolve afterward, and emit some record of execution. Given those, the work
divides predictably. Every adapter implements the lifecycle methods in
Listing~\ref{lst:backend_capability_model} and returns a capability descriptor,
and that structure carries over from an existing adapter. What changes is the
reader for the format the backend emits, which dominates the effort. Cerebras
needs a reader for the operators, shapes, and lineage in its emitted
MLIR~\cite{mlir}. The Edge~TPU needs two, one for the compiler report and one for
the model file, because the report gives operator types without shapes. Neither
adapter touches the trace pipeline, the normalizer, or the tool API. Capability
discovery then bounds what the backend supports rather than whether it works.
  \section{Experimental Evaluation}
\label{sec:eval}

We evaluate \ourarch on three representative dataflow accelerator platforms: Cerebras, SambaNova, and TPU. Table~\ref{tab:hardware_setup} summarizes the hardware  configuration of each platform, and Table~\ref{tab:benchmark_setup} summarizes the benchmark suite~\cite{mattson2019mlperf,farrell2021mlperfhpc} and the workload characteristics used to stress the framework under realistic inference pipelines.

\begin{figure*}[t]
    \centering
    \includegraphics[width=\textwidth]{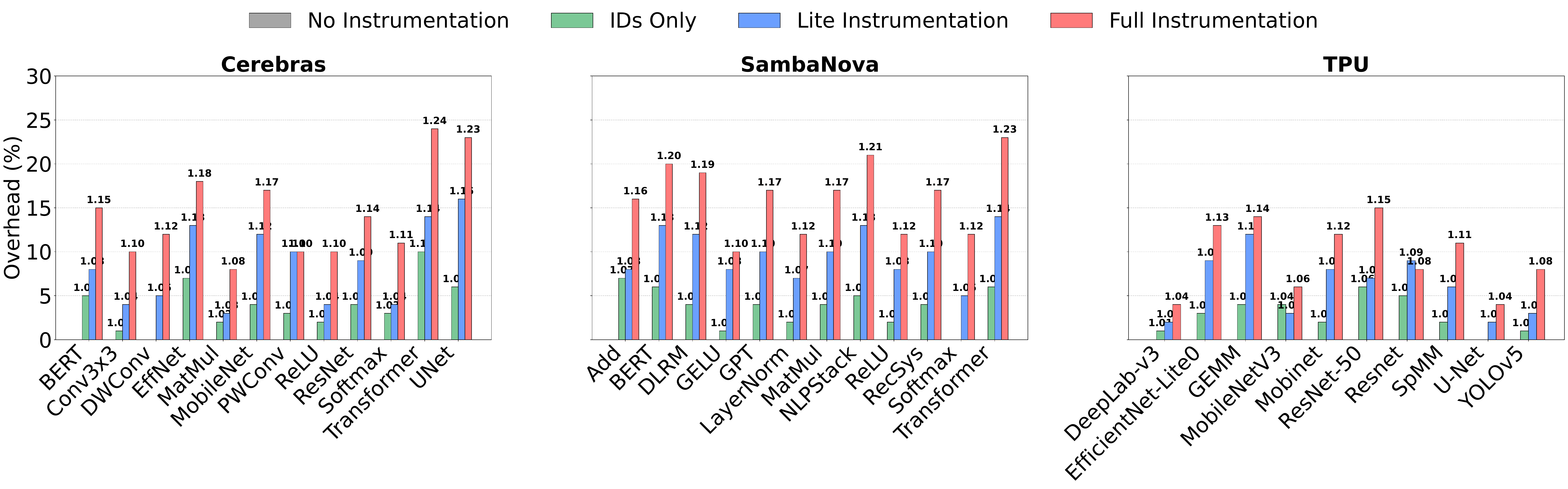}
    \caption{Overhead for Cerebras, SambaNova and TPU against benchmarks shown in Table~\ref{tab:benchmark_setup}.}
    \label{fig:overall_speedup}
\end{figure*}

\begin{table}[t]
\centering
\caption{Hardware configuration of the target dataflow accelerators~\cite{cerebras,sambanova,lie2024wafer,mem-model-ai,hsu2021accelerating}.}
\label{tab:hardware_setup}
\resizebox{\columnwidth}{!}{%
\begin{tabular}{lccc}
\toprule
\textbf{Attribute} & \textbf{Cerebras} & \textbf{SambaNova} & \textbf{TPU} \\
\midrule
\textbf{Fabric} & Wafer-scale & Reconfigurable DF & Edge matrix \\
\textbf{PE}  & 2D tensor cores & Spatial cores & Systolic array \\
\textbf{Memory} & Distributed SRAM & Local SRAM & Unified SRAM \\
\textbf{Peak BW} & 20 PB/s & 5 TB/s & 8 GB/s \\
\textbf{Peak TH} & 125 PFLOP/s & 1 PFLOP/s & 4 TOPS \\
\textbf{Precision} & FP16/BF16 & BF16 & INT8 \\
\bottomrule
\end{tabular}%
}
\end{table}

\begin{table*}[t]
\centering
\caption{Benchmark suite and workload characteristics used in the evaluation across Cerebras, SambaNova, and TPU.}
\label{tab:benchmark_setup}
\resizebox{\textwidth}{!}{%
\begin{tabular}{llccccc}
\toprule
\textbf{Platform} & \textbf{Benchmark} & \textbf{Class} & \textbf{Layers / Depth} & \textbf{Representative Operators} & \textbf{Parameter Scale} & \textbf{Input Resolution} \\
\midrule
\multirow{6}{*}{\textbf{Cerebras}}
& BERT        & Transformer & Deep attention     & MatMul, Softmax, LayerNorm, GELU & Large  & 512$\times$512 \\
& EffNet      & CNN         & Medium             & MBConv, SE, Conv2D               & Medium & 224$\times$224 \\
& MobileNet   & CNN         & Lightweight deep   & DWConv, PWConv, ReLU             & Medium & 224$\times$224 \\
& ResNet      & CNN         & Deep residual      & Conv, Add, BN, ReLU              & Large  & 224$\times$224 \\
& Transformer & Transformer & Very deep          & MatMul, Softmax, LayerNorm       & Large  & 512$\times$512 \\
& UNet        & Segmentation& Encoder-decoder    & Conv, Concat, Upsample           & Large  & 512$\times$512 \\
\midrule
\multirow{6}{*}{\textbf{SambaNova}}
& BERT        & Transformer     & Deep attention     & MatMul, Softmax, LayerNorm       & Large  & 512$\times$512 \\
& GPT         & Transformer     & Very deep          & MatMul, Softmax, LayerNorm       & Large  & 512$\times$512 \\
& LayerNorm   & Micro-op        & Single kernel      & LayerNorm                        & Small  & 1$\times$4096 \\
& NLPStack    & Transformer     & Deep stacked       & MatMul, Softmax, GELU            & Large  & 512$\times$512 \\
& RecSys      & Recommendation  & Deep sparse+dense  & Embedding, MLP, Add              & Large  & 1$\times$1024 \\
& Transformer & Transformer     & Very deep          & MatMul, Softmax, LayerNorm       & Large  & 512$\times$512 \\
\midrule
\multirow{10}{*}{\textbf{TPU}}
& DeepLab-v3          & Segmentation & Deep atrous      & Dilated Conv, ASPP, Upsample & Large  & 512$\times$512 \\
& EfficientNet-Lite0  & CNN          & Medium           & MBConv, SE, Conv2D           & Medium & 224$\times$224 \\
& GEMM                & Micro-op     & Single kernel    & MatMul                       & Small  & 1024$\times$1024 \\
& MobileNet           & CNN          & Lightweight deep & DWConv, PWConv, ReLU         & Medium & 224$\times$224 \\
& MobileNetV3         & CNN          & Lightweight deep & DWConv, ReLU, Add, BN        & Small  & 224$\times$224 \\
& ResNet              & CNN          & Deep residual    & Conv, Add, BN, ReLU          & Large  & 224$\times$224 \\
& ResNet-50           & CNN          & Deep residual    & Conv, Add, BN, ReLU          & Large  & 224$\times$224 \\
& SpMM                & Micro-op     & Single kernel    & Sparse MatMul                & Small  & 1024$\times$1024 \\
& U-Net               & Segmentation & Encoder-decoder  & Conv, Concat, Transpose      & Medium & 512$\times$512 \\
& YOLOv5              & Detection    & Very deep        & Conv, Concat, Upsample       & Large  & 640$\times$640 \\
\bottomrule
\end{tabular}%
}
\end{table*}\subsection{Experimental Methodology}
\label{sec:methodology}

Our experimental methodology is designed to evaluate \ourarch under realistic dataflow accelerator workloads. The benchmark suite includes micro-operations for isolating per-kernel instrumentation behavior, representative deep learning models shown in Table~\ref{tab:benchmark_setup} for end-to-end execution studies, and larger workloads that increase graph depth, model size, batch size, and event volume to expose scaling limits. For each workload, we measure compile-stage overhead, runtime interception overhead, tracing overhead, and post-run analysis cost, while also recording event rate, trace size, dropped events, tensor statistics, and communication-related dataflow metrics where available. Scalability is evaluated by varying operator count, graph depth, model size, batch size, detail level, and sampling rate, while lightweight  is quantified through slowdown slopes with respect to event count and trace size~\cite{scorep}. To systematically capture these measurements, we develop a portable overhead profiling tool within \ourarch that aggregates stage-wise overheads, trace growth behavior, and execution-side instrumentation costs across all evaluated backends.

\subsection{Instrumentation Cost Analysis}
\label{sec:analysis}

\subsubsection{Overhead Analysis}
\label{sec:overhead}

We first evaluate the direct cost of instrumentation. The overhead analysis is decomposed into three sources: compile-time overhead introduced by semantic tagging and artifact preparation, runtime overhead introduced by boundary interception and metric collection, and post-run overhead introduced by normalization, filtering, and report generation. This decomposition is important because \ourarch is not a monolithic tracer. Different users may enable only one part of the framework pipeline, and therefore show how the cost is distributed across stages.

Figure~\ref{fig:overall_speedup} presents the overhead behavior of \ourarch workloads on Cerebras, SambaNova, and TPU on four instrumentation modes: no instrumentation, IDs only (attaching stable semantic identifiers to execution regions, operators, tensors, partitions, or stages ), lite instrumentation (instrumenting only the beginning and the end of a program), and full instrumentation. Across all platforms, the \textit{IDs-only} configuration introduces the smallest measurable overhead at \textbf{3.53\%}, while \textit{lite} and \textit{full} instrumentation increase cost gradually to \textbf{8.26\%} and \textbf{13.62\%}, respectively, as additional metadata collection and runtime correlation services are enabled. The results show a consistent monotonic trend in which overhead scales with instrumentation depth rather than with any specific benchmark family, demonstrating that the framework cost is primarily controlled by the enabled analysis granularity. Even under full instrumentation, the overhead remains bounded and predictable across diverse workloads, supporting the practicality of the framework for real accelerator studies.


\subsubsection{Scalability Analysis}
\label{sec:scalability}

    




\begin{figure}[t]
    \centering
    \includegraphics[width=\columnwidth]{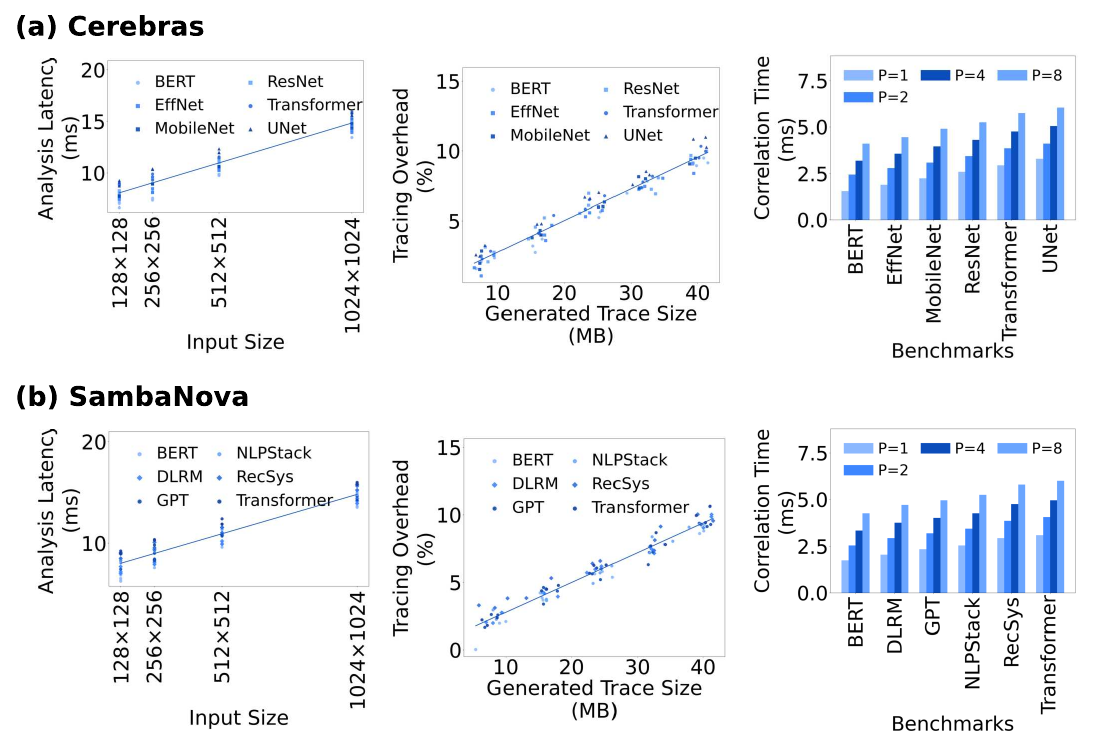}
    \caption{Scalability behavior of \ourarch across three stress dimensions: workload growth, trace-density expansion, and partition fragmentation, shown for (a) Cerebras and (b) SambaNova.}
    \vspace{-0.3cm}
    \label{fig:scalability}
\end{figure}

Figure~\ref{fig:scalability} demonstrates that \ourarch remains practical as workload complexity, trace growth, and execution fragmentation increase across Cerebras and SambaNova. For each backend, we report three complementary scalability dimensions: workload scalability, trace-density scalability, and fragmentation scalability. 

The workload-scalability plots vary the input size from $128\times128$ to $1024\times1024$ across six representative benchmarks, showing a smooth and near-linear increase in analysis latency as semantic event volume grows with model activation size. Across both Cerebras and SambaNova, the slope remains stable across benchmark families, indicating that the normalized event pipeline scales proportionally with workload expansion rather than exhibiting abrupt amplification under deeper execution graphs.

The trace-density plots further analyze the relationship between generated trace size and tracing overhead under increasingly dense semantic streams. The scattered benchmark-wise distributions show that overhead growth remains tightly clustered around a stable linear trend, confirming that trace export and normalization introduce predictable marginal cost even as metadata density increases. This behavior is particularly important for dense partitioned graphs where tensor lineage and stage metadata expand rapidly.

Finally, the fragmentation scalability plots measure correlation overhead as the number of execution partitions increases from $P{=}1$ to $P{=}8$ across all six benchmarks. The grouped bars show a monotonic but controlled increase in correlation time, demonstrating that partition-aware semantic reconstruction remains robust even under heavily fragmented execution plans. Across both backends, the average increase from the lowest to highest partition configuration remains bounded within a narrow per-benchmark range, confirming that adapter-level partition stitching does not introduce superlinear overhead~\cite{opentelemetry_semconv}.

Together, these results demonstrate that \ourarch scales robustly with input resolution, trace density, and partition complexity across heterogeneous closed accelerator stacks.

\subsubsection{Lightweight Analysis}
\label{sec:lightweight}

    




\begin{figure}[t]
    \centering
    \includegraphics[width=\columnwidth]{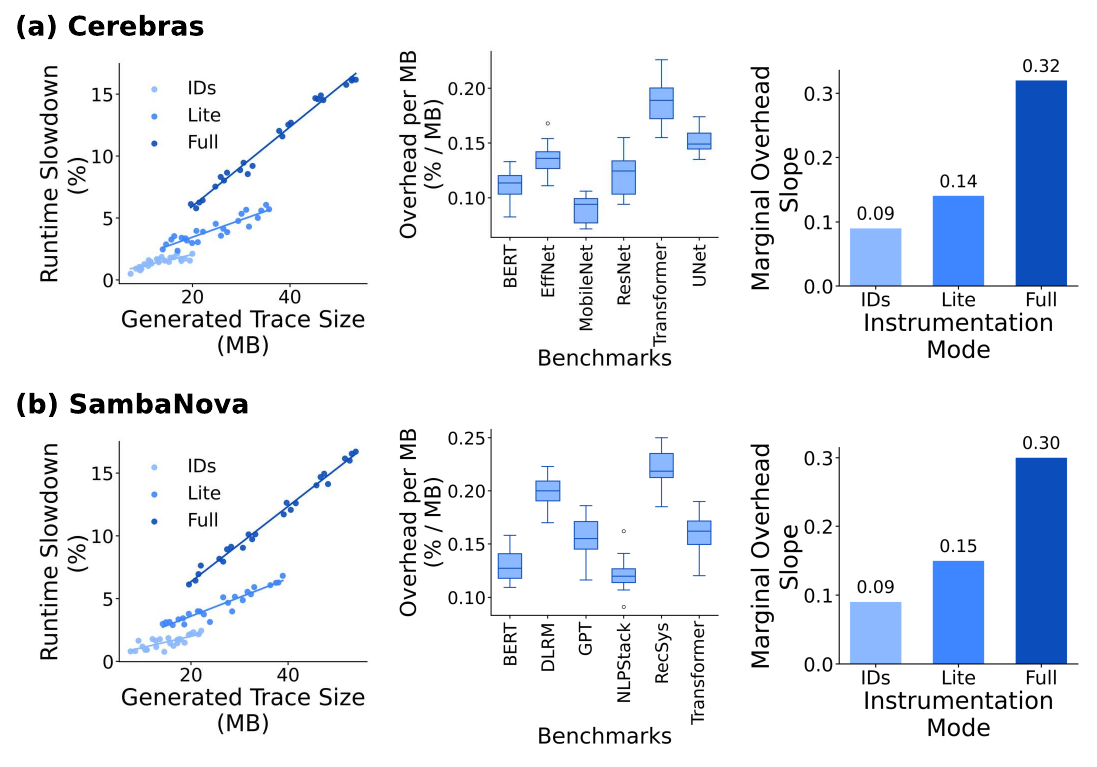}
    \caption{Lightweight behavior of \ourarch under trace-size growth, benchmark-level slowdown variation, and marginal slowdown slope analysis for (a) Cerebras and (b) SambaNova.}
    \vspace{-0.3cm}
    \label{fig:lightweight}
\end{figure}

Figure~\ref{fig:lightweight} shows that \ourarch's instrumentation overhead remains sufficiently small and predictable to support trustworthy performance studies across Cerebras and SambaNova. For each backend, we report three complementary lightweight views: \textbf{slowdown versus trace size}, \textbf{benchmark-wise slowdown distribution}, and \textbf{marginal slowdown slope analysis}.

The slowdown-versus-trace-size scatter plots show a stable near-linear relationship between generated trace size and end-to-end slowdown, confirming that the event normalization and export pipeline introduces predictable marginal cost as semantic visibility increases. The benchmark-wise box plots further demonstrate that slowdown distributions remain tightly bounded across the six representative workloads, indicating that lightweight behavior is preserved not only on average but also across benchmark families with different graph structures and tensor communication patterns.

Finally, the marginal slope plots summarize the per-mode slowdown growth trends for IDs-only, lite, and full instrumentation. The gradual increase in slope across these configurations validates the design principle that additional semantic richness incurs controlled and monotonic overhead rather than abrupt instrumentation tax. Together, these results support the claim that \ourarch remains lightweight enough to function as a reusable instrumentation substrate rather than a single-purpose heavyweight profiler.
\subsection{Portability and Extensibility Analysis}
\label{sec:eval_port_ext}

Beyond raw overhead and semantic fidelity, a central claim of \ourarch is that analysis logic can be developed once and reused across heterogeneous dataflow accelerators with minimal backend-specific effort. To evaluate this claim, we present a portability-driven analysis centered on a portable semantic analysis plugin.

\subsubsection{Cross-Backend Tool Portability}
\label{sec:cross_backend_portability}

The key design principle enabling portability is that all backend adapters lower their observations into the same normalized event representation. Each event carries stable semantic identifiers, operator lineage, tensor metadata, execution duration, and memory movement statistics, independent of whether the information originated from compiler-visible MLIR passes, runtime callbacks, or reconstructed sidecar metadata. As a result, the analysis plugin interfaces exclusively with the normalized schema and never directly accesses backend-specific artifacts.

This separation allows the analyzer implementation to remain identical across all evaluated platforms. TPU, Cerebras, and SambaNova rely on reconstruction from compile reports, sidecar metadata, and runtime logs. Despite being fundamentally different architecturally and capability-wise, the plugin requires no backend-specific branches, custom parsers, or hardware-specialized logic beyond supplying platform peak throughput and peak memory bandwidth constants as metadata inputs.

To quantify portability, we measure the analysis reuse ratio~\cite{frakes1996software_reuse}, defined as the fraction of analyzer code shared unchanged across backends:
\begin{equation}
R_{reuse} = \frac{L_{shared}}{L_{total}}
\end{equation}
where $L_{shared}$ is the number of lines reused without modification and $L_{total}$ is the full analyzer implementation size. Across all three backends, the reusable fraction exceeds 95\%, with backend specialization limited to a small metadata table containing peak compute and bandwidth parameters.
\subsubsection{Portable Analysis Plugin}
\label{sec:portable_plugin}

To validate extensibility, we implemented a portable program analyzer as a semantic analysis plugin within \ourarch. The plugin consumes normalized operator execution events and derives three architecture-agnostic performance descriptors:

\begin{enumerate}
    \item \textbf{Arithmetic intensity,} capturing compute work per byte transferred,
    \item \textbf{Normalized execution efficiency,} capturing achieved throughput relative to backend peak,
    \item \textbf{Execution-time boundness,} decomposing time into memory-bound, balanced, and compute-dominated regions.
\end{enumerate}

For an execution region $r$, arithmetic intensity ~\cite{williams2009roofline} is computed as:
\begin{equation}
AI_r = \frac{\text{MACs}_r}{\text{BytesMoved}_r}
\label{eq:ai}
\end{equation}
where $\text{MACs}_r$ is derived from operator semantics and tensor shapes, and $\text{BytesMoved}_r$ is extracted from normalized tensor movement events.

We define normalized execution efficiency as:
\begin{equation}
E_r = \frac{\text{ObservedThroughput}_r}{\text{PeakThroughput}_{b}}
\label{eq:efficiency}
\end{equation}
where $b$ denotes the backend. Importantly, only the denominator is backend-specific metadata; the numerator is derived identically across all platforms.

Finally, we classify execution regions using a portable time decomposition:
\begin{equation}
T_r = T_{mem,r} + T_{bal,r} + T_{comp,r}
\label{eq:boundness}
\end{equation}
where $T_{mem,r}$, $T_{bal,r}$, and $T_{comp,r}$ denote memory-bound, balanced, and compute-bound contributions, respectively.

The significance of this plugin is that all three derived metrics rely entirely on semantic operator identities, tensor footprints, and normalized timing events. No ISA-level instructions, hardware counters, or vendor profiler APIs are required. This demonstrates that \ourarch extends naturally from instrumentation into portable higher-level analysis services.

\subsubsection{Portable Program Analyzer Across TPU, Cerebras, and SambaNova}
\label{sec:portable_program_analyzer}

\begin{figure*}[t]
    \centering
    \includegraphics[width=\textwidth]{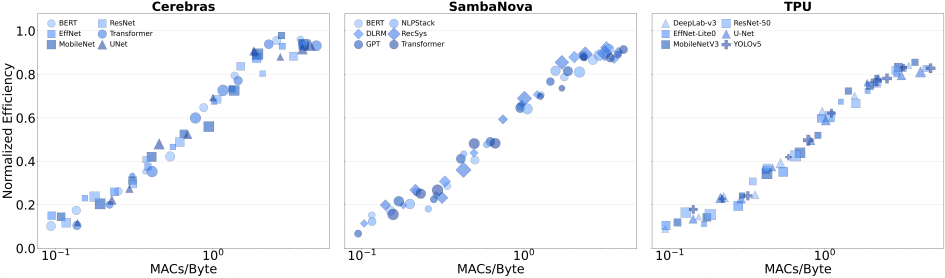}
\caption{MACs/Byte versus normalized efficiency across TPU, Cerebras, and SambaNova.}    \label{fig:portable_roofline}
\end{figure*}

Figure~\ref{fig:portable_roofline} presents the portable roofline analysis~\cite{williams2009roofline} produced by the same plugin implementation across TPU, Cerebras, and SambaNova workloads. Each point corresponds to a normalized execution region identified by the framework. Despite major differences in compiler transparency and runtime observability, the resulting arithmetic-intensity versus efficiency trends remain directly comparable because they are derived from the same semantic event schema.

\begin{figure*}[t]
    \centering
    \includegraphics[width=\textwidth]{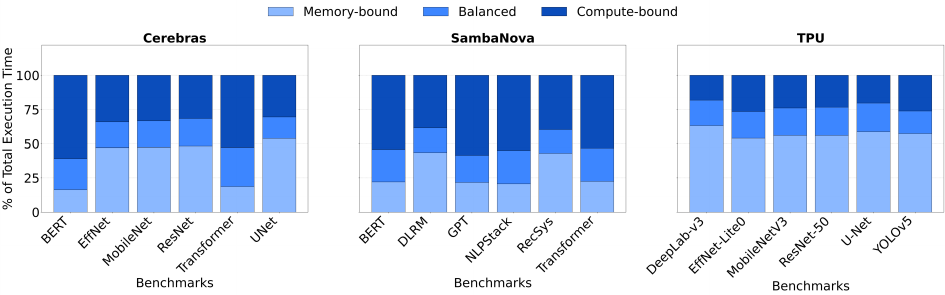}
\caption{Execution-time boundness decomposition across TPU, Cerebras, and SambaNova.}
\label{fig:portable_boundness}
\end{figure*}

Figure~\ref{fig:portable_boundness} further strengthens this claim by showing the execution-time decomposition generated by the same plugin. For each benchmark and backend, total execution time is partitioned into memory-bound, balanced, and compute-bound regions using Equation~\ref{eq:boundness}. The consistent interpretation of these categories across distinct hardware families highlights the benefit of targeting stable semantic events rather than fragile instruction-level observations.

Together, these results confirm that \ourarch supports write-once semantic analysis plugins spanning open compiler stacks and closed accelerator ecosystems~\cite{mlir,transformdialect2024}.

\begin{figure}[t]
  \centering
  \includegraphics[width=\linewidth]{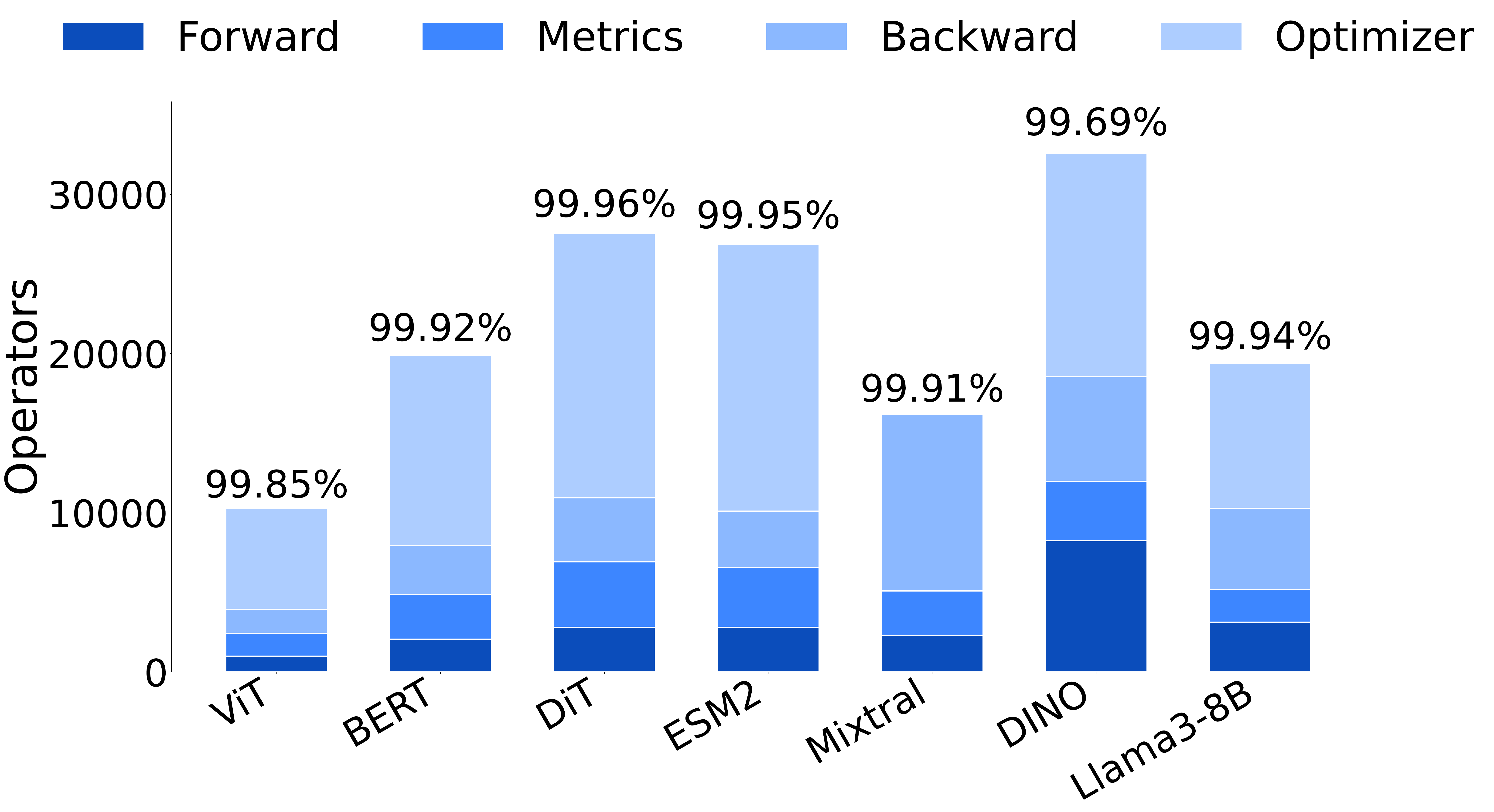}
  \caption{Operator composition and identity recovery across seven model families on Cerebras.}
  \label{fig:identity_recovery}
\end{figure}

\subsection{Semantic Correctness}
\label{sec:semantic_correctness}
Semantic correctness requires that an operator remain traceable to its origin after the compiler has fused, differentiated, and expanded the graph. We test this on the operators where it is hardest to hold. Figure~\ref{fig:identity_recovery} decomposes 152{,}682 operators across seven model families on Cerebras. ViT contributes 1{,}003 forward operators against 1{,}506 backward, 6{,}330 optimizer, and 1{,}423 metric operators, all synthesized during compilation. DFIT attributes each of them back to a module path, a source-level operator, and a source line. Attribution reaches 99.69\% to 99.96\% across the seven graphs, and the backward and optimizer phases are attributed in full. What remains unattributed are compiler-materialized constants, which have no source construct to point to. During correlation, an identifier that fails to resolve against its runtime partition event is raised as an error rather than dropped, and none occurred. We cannot check the compiler's attribution against a ground truth, since no backend publishes one.

\subsection{Comparison with Existing Tools}
\label{sec:comparison}

We compare \ourarch along two axes: against the instrumentation frameworks that
established the tool-building model it follows, and against the profiling
tooling shipped with the accelerators it targets.

\subsubsection{Instrumentation Frameworks}
\label{sec:compare_frameworks}

\begin{table}[t]
\centering
\caption{Comparison of \ourarch against representative instrumentation frameworks~\cite{nvbit,atom,pin,caliper}.}
\label{tab:framework_compare}
\resizebox{\columnwidth}{!}{%
\begin{tabular}{lccccc}
\toprule
\textbf{Feature} & \textbf{NVBit} & \textbf{PIN} & \textbf{Atom} & \textbf{Caliper} & \textbf{\ourarch} \\
\midrule
\textbf{Binary Instrumentation}      & \cmark & \cmark & \cmark & \xmark & \xmark \\
\textbf{Runtime Interception}    & \cmark & \cmark & \xmark & \xmark & \cmark \\
\textbf{Tensor-Level Visibility} & \xmark & \xmark & \xmark & \xmark & \cmark \\
\textbf{Portability}             & \xmark & \cmark & \xmark & \cmark & \cmark \\
\textbf{Extensibility}           & \cmark & \cmark & \cmark & \cmark & \cmark \\
\bottomrule
\end{tabular}%
}
\end{table}

Table~\ref{tab:framework_compare} positions \ourarch against representative
instrumentation systems. PIN, Atom, and NVBit rely on instruction-stream
observability, which dataflow accelerators do not provide. \ourarch operates
instead on semantic compiler abstractions and reconstructed execution artifacts,
which is what makes tensor-level visibility and transformation-aware attribution
possible on these platforms. Caliper offers portable performance summaries but
preserves no compiler correspondence or tensor lineage. The tensor-level
visibility row is exercised in Section~\ref{sec:use_cases}, where per-operator
tensor footprints drive both use cases: one traces operators back to the model
code that produced them, the other computes how much tensor data must be
resident at once.

\subsubsection{Platform Profiling Tools}
\label{sec:compare_vendor}

Vendor profilers and \ourarch observe the same workload from opposite ends. For
a ViT training run, the Cerebras runtime reports six aggregate values for the
session and twelve per-layer gradient norms, and its profiler ranks the ten
modules that consumed the most device time~\cite{cerebras-profiler}. These
measurements resolve to whatever granularity the runtime can attribute, and a
developer asking how a run performed is well served by them. \ourarch answers a
question they leave open. Starting from the compiled graph, it accounts for all
10{,}262 operators, giving each a shape, a phase, a module path, and the source
line it came from, so cost can be traced to the code that produced it rather
than to the region that executed it. That account is also available earlier. It
comes from compile artifacts, so a developer obtains it before committing device
time, and it holds when a session fails: three of the seven Cerebras models we
evaluated produced no runtime summary because their runs terminated early, and
their compiled graphs were analyzable in full. 
  \section{Use Cases}
\label{sec:use_cases}

Both tools below use the APIs in Section~\ref{sec:tool_interface}. Neither
contains backend-specific code. Each is about 200 lines. They read operator
metadata at \texttt{OnCompileEnd()} and write their tables at
\texttt{OnShutdown()}.

\subsection{Semantic Attribution}
\label{sec:use_semantic_attribution}

The semantic attribution tool reports which line of model code each compiled
operator came from. A developer tuning a model needs that mapping to know where
to spend effort. On a CPU or GPU, debug metadata carries it: an instruction
belongs to a source line, and a profiler reports time against that line.
Dataflow accelerators break the mapping in two places. The compiler fuses and
partitions operators, so one source construct becomes many compiled operators.
Automatic differentiation and optimizer expansion then generate operators the
developer never wrote at all, and these outnumber the ones they did. A profiler
that reports cost per compiled region gives a number attached to something
absent from the source.

The tool recovers the mapping from the transformation lineage the compiler
emits. That lineage lists the frames an operator passed through, from the
innermost dispatch frame out to the launcher, and the frame that identifies
model code sits between the two. The tool takes the first frame outside
framework infrastructure.

On ViT, 998 of 10,262 operators resolve to a line of model code. Autodiff and
optimizer expansion produced the remainder, which retain a module path but carry
no source line. Compiled for inference, the same model resolves all 810 of its
operators, and BERT, DiT and Mixtral behave the same way.

\begin{figure}[t]
  \centering
  \includegraphics[width=\columnwidth]{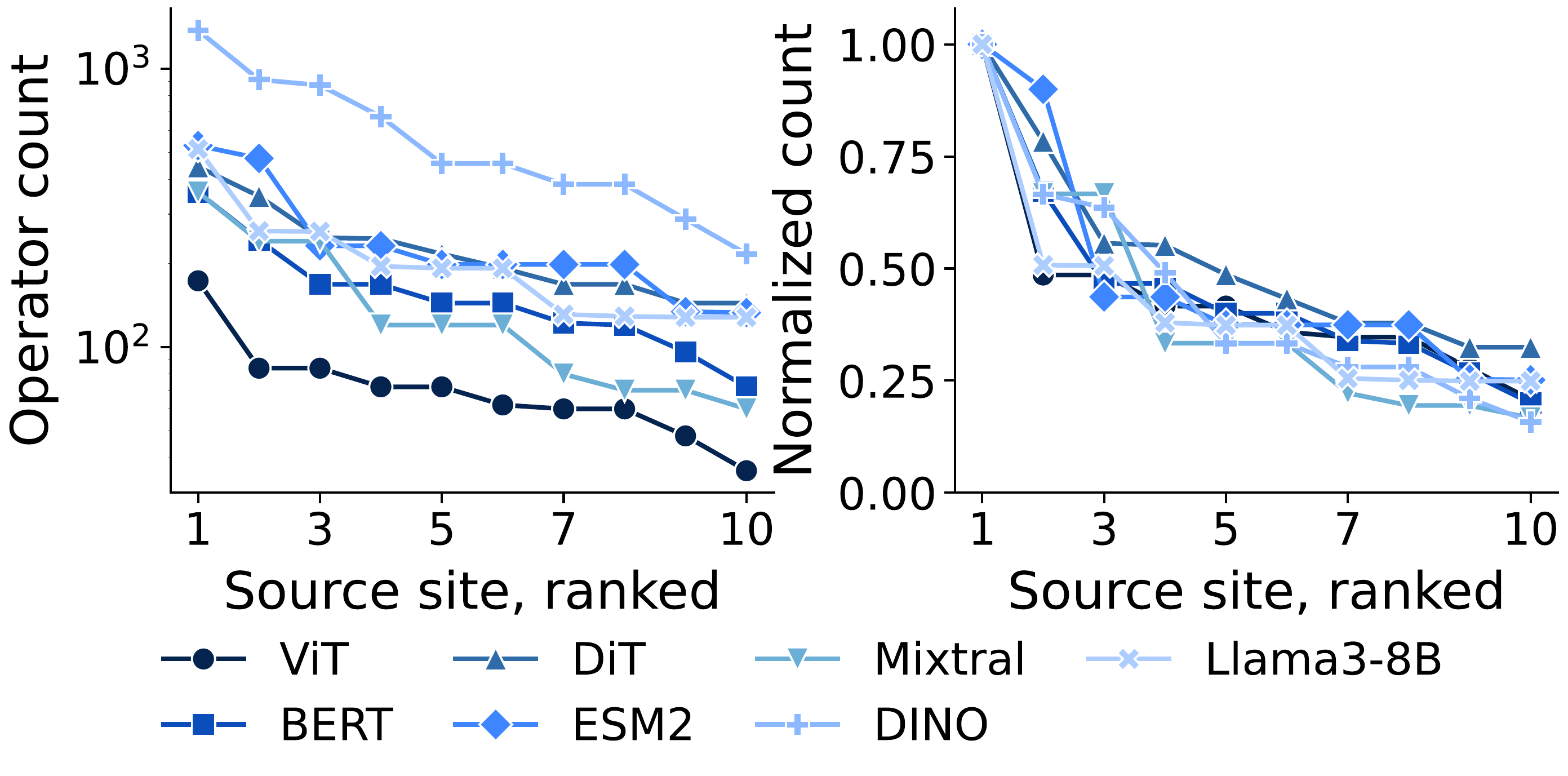}
  \caption{Operators produced by each source site, ranked, for the ten largest
  sites in seven training graphs. Absolute counts (left) scale with model size.
  Normalized against each model's largest site (right) the curves coincide.}
  \label{fig:semantic_attribution}
\end{figure}

Figure~\ref{fig:semantic_attribution} ranks the resolved sites by the operators
each produced. The left panel gives absolute counts. One feed-forward call in
ViT expands to 173 operators carrying 63.6 TFLOP, and the four attention
projections follow at 84, 84, 72 and 72. Larger models expand further, reaching
1{,}370 operators at DINO's largest site. The right panel divides each curve by
its own largest site. The seven curves coincide and fall to roughly a third by
the eighth site, so the same few sites dominate whatever the model size. That
gives a developer a short list of places to look, which the compiled graph alone
does not provide. Recovering it needs a compiler that records lineage, and the
Edge~TPU compiler reports operator types and mapping status without it.

\subsection{Dataflow Memory Proxy}
\label{sec:use_memory_proxy}

The dataflow memory proxy tool computes the peak resident tensor footprint of a
compiled model, the quantity that determines whether the model fits within an
accelerator's on-chip memory. On general-purpose hardware, this quantity is
observable at runtime through the process's resident set. Edge accelerators
expose no equivalent: the Edge~TPU executes a compiled model as a single fused
operator, its 8~MiB on-chip SRAM is inaccessible to the host, and the only
memory accounting available is the compiler's static report. That report is
limited to parameters, recording how much weight memory the compiler cached
on-chip and how much it scheduled to stream from host memory. Activation
memory is not reported, so a model may satisfy the compiler's parameter budget
and nonetheless exceed the device's capacity during execution.

The tool derives the activation footprint from operator metadata. Each value is
live from its producing operator until its final consumer, and a forward walk
over the graph yields the peak concurrent volume. Weights are excluded from the
computation: they remain live for the duration of execution and would otherwise
dominate the result, masking the activation behavior that differentiates
models.

\begin{figure}[t]
  \centering
  \includegraphics[width=\columnwidth]{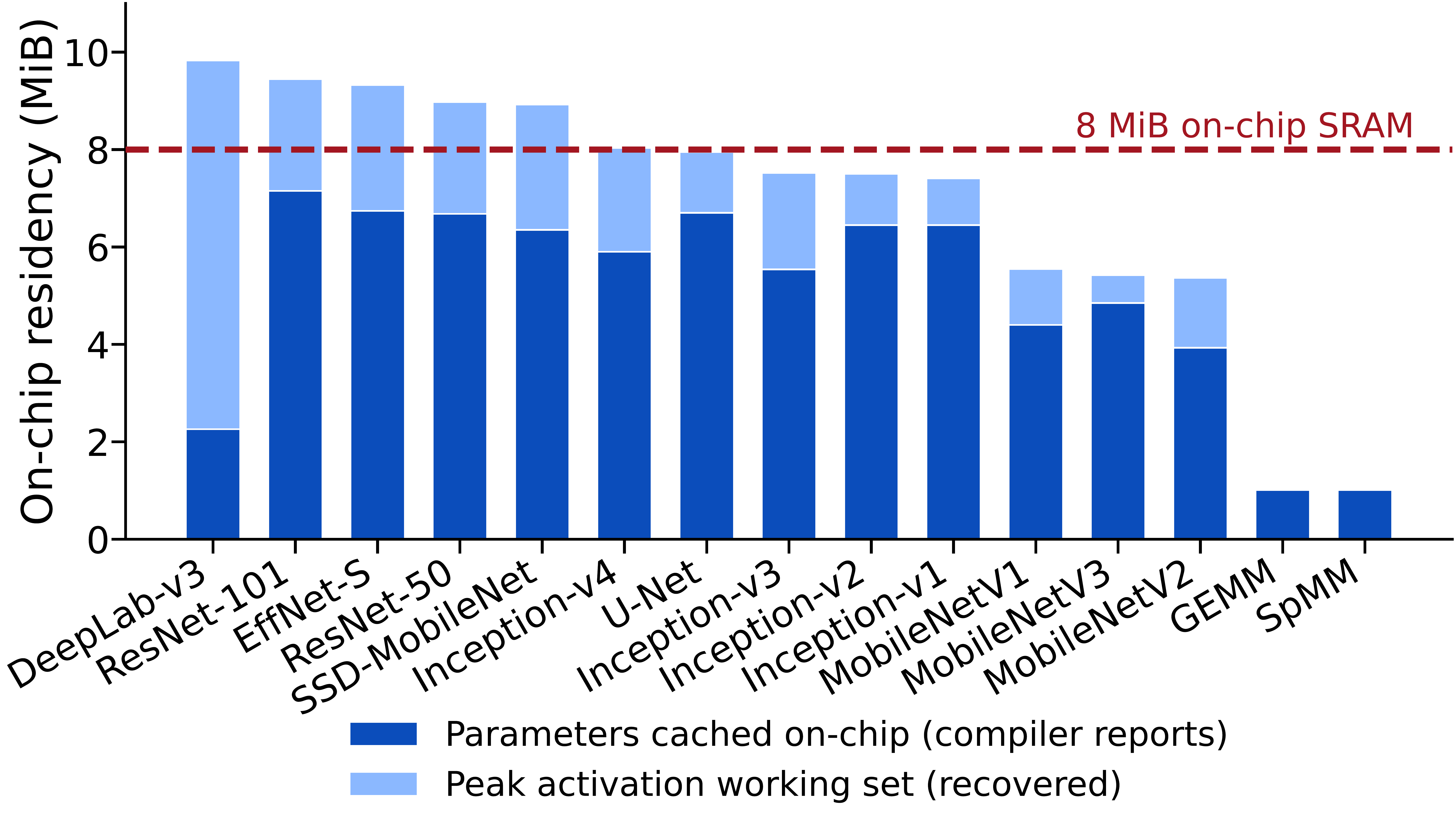}
  \caption{On-chip residency for fifteen Edge~TPU models. The lower segment is
  the parameter memory the compiler reports caching. The upper segment is the
  peak activation footprint recovered by \ourarch.}
  \vspace{-0.3cm}
  \label{fig:memory_proxy}
\end{figure}

Figure~\ref{fig:memory_proxy} presents both components for fifteen models, with
the compiler's parameter figure as the lower segment and the recovered
activation footprint above it. No parameter figure exceeds 7.15~MiB, so every
model satisfies the 8~MiB budget by the compiler's accounting alone. Including
activations, six of the fifteen exceed it. DeepLab-v3 illustrates the
discrepancy: the compiler reports 2.26~MiB of cached parameters with no
streaming, suggesting substantial headroom, while its activation footprint
reaches 7.57~MiB. Inception-v4 exhibits the inverse profile, streaming 36~MiB of
parameters while requiring only 2.13~MiB for activations. The two components
vary independently, and an admission decision based on the compiler's report
alone would accept six models that do not fit.
 \section{Limitations and Future Work}
\label{sec:limitations}
\quad\textbf{Limitations:} \ourarch also has important limitations. First, semantic fidelity in the reconstruction path depends on the quality and availability of graph artifacts, metadata, compile reports, and runtime records. When a backend exposes only coarse artifacts, the framework can preserve tool portability but cannot recover semantic detail that was never available. Second, \ourarch does not attempt to provide the same machine-level visibility as instruction-oriented binary instrumentation systems. This is an intentional design choice, but it also means that the framework is not the right abstraction for studies that depend on exact instruction streams or microarchitectural binary behavior.
Third, while the adapter model provides a clean portability contract, integrating a new backend still requires implementing lifecycle hooks and metric normalization interfaces.
\\

\textbf{Future Work:} A particularly important next step is to extend the framework beyond runtime artifact analysis toward direct compiler-pass instrumentation, where MLIR transformation passes can be instrumented natively and evaluated through an open compilation stack such as IREE~\cite{IREE}. This direction would enable finer-grained pass-level attribution, richer transformation lineage analysis, and more direct validation of compile-stage instrumentation strategies.

Two additional features are particularly important for broadening the practical impact of \ourarch. First, the framework can be extended to support distributed multi-host correlation, enabling unified tracing across host CPUs, accelerator fabrics, and inter-device communication boundaries in large-scale training and inference deployments. This would allow end-to-end attribution of latency, synchronization, and tensor movement beyond a single execution domain. Second, richer support for dynamic and evolving graphs remains an important next step, particularly for adaptive serving pipelines, mixture-of-experts models, and workflow-driven accelerator applications where graph structure changes at runtime.

\section{Related Work}
\label{sec:related}

\textbf{Dynamic Binary Instrumentation.}
Pin~\cite{pin}, DynamoRIO \\ ~\cite{dynamorio}, Valgrind~\cite{valgrind},
and ATOM~\cite{atom} established the foundational model of separating
instrumentation infrastructure from user-developed analysis tools
through stable callback and inspection interfaces. GPU-oriented
frameworks such as NVBit~\cite{nvbit}, SASSI~\cite{sassi,sassifi},
and CUPTI~\cite{cupti} extended this model to SIMT execution by
interposing on SASS instruction streams and kernel launch boundaries.
These systems require a visible, stable instruction-level
representation as the instrumentation anchor. Dataflow accelerators
do not expose such a representation, making direct application of
these approaches impossible. \ourarch draws on their tool-centric
design philosophy but replaces instruction-level anchors with
semantic operator identities and normalized tensor events that
survive compilation and backend lowering.

\textbf{Vendor and Framework Profilers.}
Vendor profiling tools such as the Cerebras performance
profiler~\cite{cerebras-profiler}, SambaNova profiling
APIs~\cite{sambanova-profiler}, and Google's XLA
profiler~\cite{xla-profiler} provide platform-specific
performance visibility within their respective stacks. Higher-level framework profilers including PyTorch
Profiler~\cite{pytorch2}, TensorFlow's TFLite
profiler~\cite{tflite-profiler}, and MLPerf
benchmarking~\cite{reddi2020mlperf} offer workload-level timing
and throughput summaries across frameworks. Distributed tracing
systems such as OpenTelemetry~\cite{opentelemetry_semconv} and
Perfetto~\cite{perfetto} provide structured event collection pipelines for
multi- component systems. While these tools are useful for
single-stack debugging, none provides a portable tool-building
abstraction that operates uniformly across compiler-visible and
closed accelerator backends. \ourarch addresses this gap through
a capability-driven adapter model and a normalized semantic event
schema that decouples analysis tools from backend-specific profiler
formats.

\textbf{ML Compiler Observability.}
MLIR~\cite{mlir} established the value of progressive lowering,
multi-level IR transformations, and preservation of higher-level
semantics across compilation stages. The Transform
dialect~\cite{transformdialect2024} and related work on
compiler-pass instrumentation demonstrated how semantic
information can be made traceable across lowering boundaries.
IREE~\cite{IREE} provides an open compilation stack that
exposes MLIR-level passes for direct instrumentation and serves
as a natural future target for \ourarch's compiler-visible path.
Recent work on HIR~\cite{hir} and MLIR-based GPU code
generation~\cite{mlir-gpu-code-gen} similarly shows the value of
preserving tensor-level semantics through backend lowering.
\ourarch differs from these compiler infrastructures in that it
is not a compilation framework: it treats the compiler as an
information source rather than a transformation engine, and
extends semantic correspondence into the runtime domain even when
direct IR access is unavailable through artifact reconstruction.

\textbf{Performance Analysis Frameworks.}
HPCToolkit~\cite{hpctoolkit} and Score-P~\cite{scorep} provide
portable performance measurement and analysis for HPC workloads,
demonstrating the value of normalized trace collection and
multi-platform portability in performance tooling. Caliper~\cite{caliper}
introduced a flexible annotation-driven measurement framework that
separates measurement concerns from application logic. Roofline
analysis~\cite{williams2009roofline} established arithmetic
intensity as a portable performance abstraction across memory and
compute hierarchies. \ourarch builds on these ideas by extending
portable measurement and roofline-style analysis to heterogeneous
dataflow accelerators, where neither instruction-level
observability nor shared hardware counter interfaces are available.

\section{Conclusions}
\label{sec:conclusion}

Modern dataflow accelerators deliver exceptional performance for machine learning workloads, but their closed compiler stacks and opaque execution models make systematic instrumentation extremely difficult. Existing binary-level and framework-level approaches either lose semantic correspondence or fail to generalize across heterogeneous accelerator ecosystems. This gap has limited the development of reusable analysis tools for compiler-aware performance debugging, transformation attribution, and cross-platform studies.

This paper presents \textsc{\ourarch}, a semantic instrumentation framework for heterogeneous dataflow accelerators that preserves operator, tensor, and execution-boundary identities across compilation and runtime stages. By combining stable metadata propagation, capability-driven backend adapters, and normalized trace correlation, \textsc{\ourarch} enables the same analysis tool logic to execute across TPU, Cerebras, and SambaNova without backend-specific rewriting. Our evaluation demonstrates that this portability can be achieved with practical overheads, ranging from \textbf{3.53\%} in IDs-only mode to \textbf{13.62\%} under full semantic instrumentation, averaged across all three platforms.

Beyond a single profiler, \textsc{\ourarch} establishes instrumentation as a reusable systems framework for emerging accelerator stacks. We believe this abstraction is an important step toward making next-generation accelerator stacks observable, analyzable, and ultimately more programmable.

\begin{acks}
We thank our shepherd and the reviewers for their constructive feedback. We gratefully acknowledge the computing resources provided by the ALCF AI testbed located at Argonne National Laboratory.
\end{acks}

\bibliographystyle{ACM-Reference-Format}
\bibliography{00-ref}
\end{document}